\documentclass[conference]{IEEEtran}
\IEEEoverridecommandlockouts

\usepackage{cite}
\usepackage{amsmath,amssymb,amsfonts}
\usepackage{amsthm}
\usepackage{graphicx}
\usepackage{textcomp}
\usepackage{balance}
\usepackage{color,xcolor}
\usepackage{multirow}
\usepackage{subfigure}
\usepackage{array}
\usepackage{enumitem}
\usepackage{epstopdf}
\usepackage{booktabs}
\usepackage{threeparttable}
\usepackage{bm}
\usepackage[colorlinks,linkcolor=blue,citecolor=blue,urlcolor=blue]{hyperref}

\newtheoremstyle{mydefn}
{}{}
{\it}  
{0pt}       
{\bfseries} 
{:}
{.5em}
{}          
\theoremstyle{mydefn}

\newtheorem{definition}{Definition}
\newtheorem{theorem}{Theorem}
\newtheorem{corollary}{Corollary}

\newtheorem{lemma}{Lemma}

\newtheoremstyle{myexample}
{}{}
{}  
{0pt}       
{\bfseries} 
{:}
{.5em}
{}          
\theoremstyle{myexample}

\renewcommand{\paragraph}[1]{\vspace{0.5em}\noindent\textbf{#1}}
\newcommand{\subparagraph}[1]{\vspace{0.5em}\noindent\textit{\underline{#1}}}

\newcommand{\norm}[1]{\Vert #1 \Vert}
\newcommand{\num}[1]{\vert #1 \vert}
\newcommand{\abso}[1]{\vert #1 \vert}
\newcommand{\ip}[1]{\langle #1 \rangle}

\makeatletter
\newif\if@restonecol
\makeatother

\usepackage[linesnumbered,ruled,vlined]{algorithm2e}
\usepackage{algpseudocode}

\SetKwComment{Comment}{$\triangleright$\ }{}

\begin{document}

\title{Lightweight-Yet-Efficient: Revitalizing Ball-Tree for Point-to-Hyperplane Nearest Neighbor Search}

\author{
\IEEEauthorblockN{Qiang Huang, Anthony K. H. Tung}

\IEEEauthorblockA{School of Computing, National University of Singapore, Singapore}

\IEEEauthorblockA{\{huangq,atung\}@comp.nus.edu.sg}
}

\maketitle

\begin{abstract}
Finding the nearest neighbor to a hyperplane (or Point-to-Hyperplane Nearest Neighbor Search, simply P2HNNS) is a new and challenging problem with applications in many research domains. While existing state-of-the-art hashing schemes (e.g., NH and FH) are able to achieve sublinear time complexity without the assumption of the data being in a unit hypersphere, they require an asymmetric transformation, which increases the data dimension from $d$ to $\Omega(d^2)$. This leads to considerable overhead for indexing and incurs significant distortion errors.

In this paper, we investigate a tree-based approach for solving P2HNNS using the classical Ball-Tree index. Compared to hashing-based methods, tree-based methods usually require roughly linear costs for construction, and they provide different kinds of approximations with excellent flexibility. A simple branch-and-bound algorithm with a novel lower bound is first developed on Ball-Tree for performing P2HNNS. Then, a new tree structure named BC-Tree, which maintains the Ball and Cone structures in the leaf nodes of Ball-Tree, is described together with two effective strategies, i.e., point-level pruning and collaborative inner product computing. BC-Tree inherits both the low construction cost and lightweight property of Ball-Tree while providing a similar or more efficient search. Experimental results over 16 real-world data sets show that Ball-Tree and BC-Tree are around 1.1$\sim$10$\times$ faster than NH and FH, and they can reduce the index size and indexing time by about 1$\sim$3 orders of magnitudes on average. The code is available at \url{https://github.com/HuangQiang/BC-Tree}.
\end{abstract}

\begin{IEEEkeywords}
Nearest Neighbor Search, Hyperplane Query, Point-to-Hyperplane Distance, Ball Tree, Cone Structure
\end{IEEEkeywords}

\section{Introduction}
\label{sect:intro}
Point-to-Hyperplane Nearest Neighbor Search (P2HNNS) plays a vital role in many research domains, such as active learning with Support Vector Machines (SVMs) \cite{schohn2000less, campbell2000query, tong2001support, vijayanarasimhan2014large}, large margin dimensionality reduction \cite{xu2014large, saberian2016large}, and maximum margin clustering \cite{xu2004maximum, zhao2008efficient, zhang2018optimal}. 
For example, in the applications of pool-based active learning with SVMs, the goal is to request labels for the data points closest (with minimum margin) to the SVM's decision hyperplane to reduce human efforts for annotation \cite{tong2001support}. 
Moreover, motivated by the success of SVM for classification, the maximum margin clustering aims at finding the hyperplane maximizing the minimum margin to the data, which can separate the data from different classes \cite{xu2004maximum, zhao2008efficient}. 
Such applications require finding the data points that are closest to the hyperplane. 

In most applications, data points are often represented as vectors in a $(d-1)$-dimensional Euclidean space $\mathbb{R}^{d-1}$, while hyperplane queries (e.g., the decision hyperplane) have a higher dimension $d$. 
For any data point $\bm{p} = (p_1,\cdots,p_{d-1})$ and hyperplane query $\bm{q} = (q_1,\cdots,q_d)$, their point-to-hyperplane (P2H) distance is defined as below: 
\begin{equation}
\label{eqn:p2h-dist}
d_{P2H}(\bm{p},\bm{q}) = \frac{\abso{q_d + \sum_{i=1}^{d-1} p_i q_i}}{\sqrt{\sum_{i=1}^{d-1} q_i^2}}.
\end{equation}

Compared with the classic point-to-point similarity search problems, such as Nearest Neighbor Search (NNS) and Furthest Neighbor Search (FNS), 
P2HNNS is a more recent and challenging problem. The reasons are two folds: 
(1) The data points and queries do not have the same dimensionality, leading to a complex computation of the P2H distance. 
(2) More importantly, unlike the commonly used distance metrics such as Euclidean and angular distances, the P2H distance is not a metric because it consists of an inner product computation and an absolute value operation, which violate the axioms of the identity of indiscernibles and the triangle inequality. 
As such, many practical and efficient similarity search methods such as Locality-Sensitive Hashing (LSH) \cite{indyk1998approximate, charikar2002similarity, datar2004locality, bawa2005lsh, andoni2006near, lv2007multi, tao2009quality, gan2012locality, sun2014srs, andoni2015practical, huang2015query, zheng2016lazylsh, lei2019sublinear, lei2020locality, lu2020r2lsh, lu2020vhp, zheng2020pm} and proximity graph \cite{dong2011efficient, malkov2014approximate, malkov2018efficient, fu2019fast, zhao2020song, guerraoui2020smaller, morozov2018non, zhou2019mobius, tan2021norm, fu2022high} cannot be directly used for the P2HNNS. 
A trivial solution for solving P2HNNS is an exhaustive scan through all points in the database, but this is usually computationally prohibitive. 

Previous researches assume that data points are all in the unit hypersphere, i.e., $\norm{\bm{p}} = 1$ for all $\bm{p}$'s. 
Based on this assumption, researchers designed several hash functions that are locality-sensitive to the angular distance between data points and the normal vector of the hyperplane query, and they proposed a series of hyperplane hashing schemes \cite{jain2010hashing, liu2012compact, vijayanarasimhan2014hashing, liu2016multilinear} for performing P2HNNS. 
Aum{\"u}ller et al. \cite{aumuller2018distance} further introduced a general distance-sensitive hashing scheme beyond LSH. 
These hashing schemes tackle the P2HNNS in sublinear time, which is significantly more efficient than the exhaustive scan, and they show great success in large-scale active learning \cite{jain2010hashing, liu2012compact, liu2016multilinear}. 
Nevertheless, the data normalization assumption might not be valid in many applications, such as clustering \cite{zhao2008efficient, zhang2018optimal} and dimension reduction \cite{xu2014large, saberian2016large}. 
For such applications, they are no longer locality-sensitive (or distance-sensitive) and degrade rapidly\cite{huang2021point}. 

Huang et al. \cite{huang2021point} proposed the first two provably asymmetric LSH schemes NH and FH for solving P2HNNS beyond the unit hypersphere. 
They appended one dimension for each data with $1$, i.e., $\bm{x} = (\bm{p};1)$, to align the dimension of data points and queries; 
then, they designed a two-step asymmetric transformation, i.e., the vector transformations $\bm{P} \circ \bm{f}: \mathbb{R}^d \rightarrow \mathbb{R}^{d(d+1)/2+1}$ and $\bm{Q} \circ \bm{g}: \mathbb{R}^d \rightarrow \mathbb{R}^{d(d+1)/2+1}$ on data points $\bm{x} \in \mathbb{R}^d$ and hyperplane queries $\bm{q} \in \mathbb{R}^d$, respectively, to convert this challenging problem into a classic NNS (or FNS) problem on Euclidean distance. 
The asymmetric transformation $\bm{P} \circ \bm{f}(\bm{x})$ and $\bm{Q} \circ \bm{g}(\bm{q})$ is the key to removing the absolute value operation and remedying the data normalization issue. 

Nonetheless, this asymmetric transformation also has two considerable limitations. 
First, it significantly increases the dimensionality of the data from $d$ to $\Omega(d^2)$. 
This increases the indexing time of both NH and FH by a multiplicative factor of $\Omega(d^2)$ \cite{huang2021point}, leading to a huge overhead for indexing. 
For example, considering a data set with a moderate dimension $d=100$, its data dimensionality after this asymmetric transformation is around $5,000$. 
Thus, compared with the methods building index directly based on the original dimension, NH and FH (with this transformation) will be slower by about $50\times$. 
Note that the query time of NH and FH also suffer from this $\Omega(d^2)$ factor \cite{huang2021point}, which might restrict their efficiency in dealing with high-dimensional data. 
Huang et al. \cite{huang2021point} suggested applying the randomized sampling strategy \cite{jain2010hashing} to approximate this transformation, which can reduce the dimensionality from $\Omega(d^2)$ to $O(\frac{1}{\epsilon^2})$, where $0<\epsilon<1$ is an estimation error. 
This strategy, however, also introduces an additive error to the hash values such that NH and FH fail to have a theoretical guarantee to deal with P2HNNS and become less promising. 

Second, with this asymmetric transformation, though the problem of P2HNNS can be converted into the well-studied problem of NNS (or FNS), it adds a large constant to the Euclidean distance $\norm{\bm{P} \circ \bm{f}(\bm{x}) - \bm{Q} \circ \bm{g}(\bm{q})}$. 
This large constant leads to a significant distortion error for the later NNS (or FNS), in the sense that the Euclidean distance between any $\bm{P} \circ \bm{f}(\bm{x})$ and $\bm{Q} \circ \bm{g}(\bm{q})$ become close to each other, i.e., given a set of data points $\mathcal{S}$, $\max_{\bm{x}\in \mathcal{S}} \norm{\bm{P} \circ \bm{f}(\bm{x}) - \bm{Q} \circ \bm{g}(\bm{q})} / \min_{\bm{x}\in \mathcal{S}} \norm{\bm{P} \circ \bm{f}(\bm{x}) - \bm{Q} \circ \bm{g}(\bm{q})} \rightarrow 1$. 
Suppose this ratio is less than 2, and we set up an approximation ratio $c=2$ for approximate NNS (which is a typical setting for LSH schemes \cite{datar2004locality, lv2007multi, tao2009quality, gan2012locality, huang2015query, huang2017query}); then, any $\bm{P} \circ \bm{f}(\bm{x})$ can be the approximate nearest (or furthest) neighbor of $\bm{Q} \circ \bm{g}(\bm{q})$ even though their P2H distance is very large, which means that the results of NH and FH can be arbitrarily bad. 

To avoid the issues of hashing-based methods, in this paper, we will look at solving the P2HNNS problem by using space partition methods, or more specifically, tree-based methods \cite{bentley1975multidimensional, guttman1984r, omohundro1989five, katayama1997sr, beygelzimer2006cover, dasgupta2008random, dhesi2010random, dasgupta2013randomized, ram2019revisiting}. 
Compared with hashing-based methods (especially LSH) and proximity graph-based methods, tree-based methods have numerous advantages. 
First, they usually require roughly linear time and space for construction, such as KD-Tree \cite{bentley1975multidimensional}, Ball-Tree \cite{omohundro1989five}, and Randomized Partition Trees \cite{dasgupta2008random, dasgupta2013randomized}. As a comparison, LSH schemes often need subquadratic time and space to build hash tables; proximity graph-based methods use roughly linear space to store their graph structures, but they usually require subquadratic time (or even quadratic time) for indexing. 
Second, tree-based methods can be adapted to many approximate cases with theoretical guarantee, including distance approximation \cite{arya1998optimal, ram2019revisiting} and rank approximation \cite{ram2009rank}. In contrast, LSH schemes only provide a distance approximation guarantee, while there exists a gap between the theory and practice for proximity graph-based methods with fewer investigations \cite{prokhorenkova2020graph}. 
Third, they provide flexibility with a limited accuracy and/or time budget, i.e., users can set up different leaf sizes and candidate fractions for different search requirements. LSH schemes usually fail if we do not check sufficient candidates, and likewise for proximity graph-based methods if we stop before they converge. 
Even though tree-based methods suffer from the \emph{curse of dimensionality} for exact queries \cite{weber1998quantitative, beyer1999nearest}, their approximate versions have great potential to deal with high-dimensional similarity search \cite{liu2004investigation, sinha2014lsh, ram2019revisiting}. 

\paragraph{Contributions}
In this paper, we study the vanilla Ball-Tree index to tackle the P2HNNS. 
Recall that the P2H distance $d_{P2H}$ is not metric and contains an absolute value operation. 
Thus, even though there exist lower bounds based on the Ball-Tree structure for the NNS on Euclidean distance \cite{omohundro1989five} and upper bound(s) for the Maximum Inner Product Search (MIPS) \cite{ram2012maximum}, they are not applicable to $d_{P2H}$. 
Motivated by this observation, we first design a new lower bound based on the Ball-Tree structure for $d_{P2H}$ and develop a simple yet efficient branch-and-bound algorithm for solving P2HNNS. 

Moreover, we propose a new tree structure named BC-Tree, which is built upon Ball-Tree while maintaining its \underline{B}all and \underline{C}one structures in the leaf nodes. 
Using the two structures, we introduce two novel lower bounds for data points and perform point-level pruning in the leaves to reduce the total candidate verification cost. 
Further, by leveraging the linear properties of the center computation and the inner product computation, we develop a collaborative inner product computing strategy for BC-Tree to cut down the total lower bound computation cost. 
We demonstrate that BC-Tree inherits both the low construction cost and lightweight property of Ball-Tree while providing a similar or more efficient search.  

We conduct a comprehensive comparison of Ball-Tree and BC-Tree with two state-of-the-art hashing schemes, NH and FH. Extensive results over 16 real-world data sets 
show that Ball-Tree and BC-Tree can reduce the indexing overhead by about 1$\sim$3 orders of magnitudes on average, and meanwhile, they are around 1.1$\sim$10$\times$ faster than NH and FH. 

\paragraph{Organization}
The roadmap of this paper is as follows. 
Section \ref{sect:settings} discusses the problem settings. 
We present Ball-Tree and BC-Tree for P2HNNS in Sections \ref{sect:ball} and \ref{sect:bc}, respectively. 
Experimental results are analyzed in Section \ref{sect:expt}. 
Section \ref{sect:related_work} surveys related work. We conclude our work in Section \ref{sect:conclusions}.

\section{Problem Settings}
\label{sect:settings}

As the data points and hyperplane queries have different dimensionality, we formalize the problem of P2HNNS with some simplifications before presenting our methods. 

\begin{table}[t]
\centering
\renewcommand{\arraystretch}{1.0}
\caption{The summary of commonly used notations in this paper.}
\label{tab:notations}
\vspace{-0.5em}
\begin{tabular}{ll} \toprule
Notations & Description \\ \midrule
$\mathcal{S}$, $n$, $d$ & a database $\mathcal{S}$ of $n$ data points in $\mathbb{R}^d$, i.e., $n = \num{\mathcal{S}}$ \\
$k$ & $k$ value in the top-$k$ P2HNNS results\\ 
$\bm{x}, \bm{q}$ & data point, query point \\ 
$\ip{\cdot,\cdot}$ & the inner product of two points \\ 
$\norm{\cdot}$  & $l_2$ norm of a point (or Euclidean distance of two points) \\ 
$N$ & a node (internal node or leaf node) of a tree structure \\ 
$N.S$  & the set of data points in a node \\ 
$N.lc$, $N.rc$ & the left child and right child of a node \\ 
$N.\bm{c}$, $N.r$  & the center and the radius of a node \\ 
$N_0$  & maximum leaf size, i.e., maximum \# points in a leaf \\ 
$\theta$, $\varphi$  & the angle of two points \\ 
$\bm{q}.bm$, $\bm{q}.\lambda$ & the current best match data and the minimum $\abso{\ip{\bm{x},\bm{q}}}$ \\ 
\bottomrule 
\end{tabular}
\vspace{-0.75em}
\end{table}

First, we append one dimension for each data point $\bm{p} \in \mathbb{R}^{d-1}$ with $1$, i.e., $\bm{x} = (\bm{p};1)=(p_1,\cdots,p_{d-1},1) \in \mathbb{R}^d$, where $(;)$ 
represents the concatenation of dimensions. Note that with this step, the numerator of Equation \ref{eqn:p2h-dist} can be reduced to an absolute inner product computation.  
Second, we assume $\sqrt{\sum_{i=1}^{d-1} q_i^2}=1$ as this term is fixed and is not equal to 0 for a certain non-trivial query $\bm{q} \in \mathbb{R}^d$; otherwise, we rescale $\bm{q}$ to satisfy this assumption, which can achieve the same P2HNNS results. 
Let $\ip{\bm{x},\bm{q}} = \sum_{i=1}^{d} x_i q_i$ be the inner product of $\bm{x}$ and $\bm{q}$. Then, the P2H distance can be simplified as below:
\begin{equation}
\label{eqn:p2h-dist-simplify}
d_{P2H}(\bm{p},\bm{q}) = \abso{\ip{\bm{x},\bm{q}}}
\end{equation}

Suppose $\mathcal{D}$ is the original data set and $\mathcal{S}$ denotes a set of data points after dimension appending, i.e., $\mathcal{S} = \{\bm{x} = (\bm{p};1) \mid \bm{p} \in \mathcal{D}\}$. 
Since $\arg\min_{\bm{p} \in \mathcal{D}} d_{P2H}(\bm{p},\bm{q}) \Leftrightarrow \arg\min_{\bm{x} \in \mathcal{S}} \abso{\ip{\bm{x},\bm{q}}}$, the problem of P2HNNS can be formalized as follows:
\begin{definition}[P2HNNS]
\label{def:p2hnns-problem}
Given a set $\mathcal{S}$ of $n$ data points in $\mathbb{R}^{d}$,  the problem of P2HNNS is to construct a data structure which, for any query point $\bm{q} \in \mathbb{R}^{d}$, finds the data point $\bm{x^*} \in \mathcal{S}$ such that the P2H distance of $\bm{x^*}$ and $\bm{q}$ is minimized, i.e.,
\begin{equation}
\label{eqn:p2h-nns-simplify}
\bm{x^*} = \arg\min_{\bm{x} \in \mathcal{S}} \abso{\ip{\bm{x},\bm{q}}}.
\end{equation}
\end{definition}

Hereafter, we use $\mathcal{S}$ to denote the data set and regard both data $\bm{x} \in \mathcal{S}$ and queries $\bm{q}$ as vectors (or points) in the same dimension, i.e., $\bm{x},\bm{q}\in \mathbb{R}^{d}$. The commonly used notations throughout this paper are summarized in Table \ref{tab:notations}.

\section{Ball-Tree}
\label{sect:ball}

\subsection{Why Considering Ball-Tree?}
\label{sect:ball:why}
Ball-Tree is a classical tree structure in the literature for many similarity search tasks \cite{omohundro1989five, samet2006foundations, ram2012maximum}. Compared with other popular tree structures such as KD-Tree \cite{bentley1975multidimensional, ram2019revisiting}, R-Trees \cite{guttman1984r, sellis1987r+, beckmann1990r}, Cover-Tree \cite{beygelzimer2006cover}, and Randomized Partition Trees \cite{dasgupta2008random, dasgupta2013randomized} and the commonly used space-filing curves such as Z-order curve \cite{tao2009quality} and Hilbert curve \cite{kamel1993packing,arora2018hd}, we choose Ball-Tree based on the following four concerns. 
\begin{enumerate}[leftmargin=28pt,label*=(\arabic*)]
\item The data structure of Ball-Tree itself is simple yet lightweight, i.e., it only requires a center and a radius to maintain a ball. As such, Ball-Tree is extremely fast to construct, which takes roughly linear time only. 

\item For other tree-based methods, such as KD-Tree and R-Tree, they usually maintain a bounding box to provide a lower bound for a specific distance (e.g., Euclidean distance). 
Nonetheless, as the P2H distance contains an absolute value operation, their lower bounds might contain $O(d)$ cases as the vertex of the bounding box in each dimension can be either at least 0 or smaller than 0, leading to a complex computation. 
In contrast, due to the simple ball structure, as will be shown in Theorem \ref{theorem:node_lower_bound}, the lower bound of Ball-Tree contains three cases only, which might be simpler to compute and analyze. 

\item For space-filling curves, their fundamental property ensures that if two points are close in the one-dimensional order, they are probably also close in the original high-dimensional space \cite{sagan2012space, bhattacharya2014fundamentals, arora2018hd}. However, since the P2H distance is not a metric, the close points in the one-dimensional order are probably not the answers to the hyperplane query. As it might be hard to design a promising space-filling curve in a non-metric space, we first look at solving the P2HNNS using Ball-Tree. 

\item With the simple yet lightweight structure, Ball-Tree is easy to combine with other optimizations for acceleration (e.g., we develop some optimizations in Section \ref{sect:bc}). Moreover, as it is a space partition method, we can leverage it to split massive data sets into fine granularities for scalable and distributed P2HNNS.
\end{enumerate}

Before we illustrate how to leverage Ball-Tree for performing P2HNNS, we first revisit its data structure and the process of constructing a Ball-Tree.

\subsection{Ball-Tree Construction}
\label{sect:ball:construct}
\vspace{-0.5em}
\paragraph{Ball-Tree Structure}
Ball-Tree is a binary space partition tree. Each node $N$ consists of a subset of data points, i.e., $N.S \subset \mathcal{S}$. Let $\num{N}$ be the number of data points in a node $N$, i.e., $\num{N} = \num{N.S}$. Any node $N$ and its two children $N.lc$ and $N.rc$ satisfy the following two properties:
\begin{align}
\num{N.lc} + \num{N.rc} &= \num{N}, \label{eqn:ball_union} \\
N.lc \cap N.rc &= \emptyset. \label{eqn:ball_overlap}
\end{align}
Specifically, $N.S = \mathcal{S}$ if $N$ is the root of Ball-Tree. Each internal (and leaf) node maintains a ball for its data points, where the center $N.\bm{c}$ and the radius $N.r$ are defined as below:
\begin{align}
N.\bm{c} &= \frac{1}{\num{N}} \sum_{\bm{x} \in N.S} \bm{x}, \label{eqn:center} \\
N.r &= \max_{\bm{x} \in N.S} \norm{\bm{x} - N.\bm{c}}. \label{eqn:radius}
\end{align}

According to Equations \ref{eqn:center} and \ref{eqn:radius}, the center $N.\bm{c}$ is the centroid of all data points $\bm{x} \in N.S$, while the radius $N.r$ is the maximum Euclidean distance between the center $N.\bm{c}$ and the data points $x \in N.S$. With $N.\bm{c}$ and $N.r$, each node $N$ can enclose its data points within a virtual ball. 

\begin{algorithm}[t]
\caption{\textsf{BallTreeConstruct}}
\label{alg:ball_construct}
\KwIn{subset $S \subset \mathcal{S}$, maximum leaf size $N_0$;}
$N.S \leftarrow S$\;
$N.\bm{c} = \frac{1}{\num{N}} \sum_{\bm{x} \in N.S} \bm{x}$\;
$N.r = \max_{\bm{x} \in N.S} \norm{\bm{x}-N.\bm{c}}$\;
\If(\Comment*[f]{leaf node}){$\num{N} \leq N_0$} {
	\Return $N$\;
}
\Else(\Comment*[f]{internal node}){
	$\bm{x_l}, \bm{x_r} \leftarrow $~\textsf{Split}($S$)\;
	$S_l \leftarrow \{\bm{x} \in S \mid \norm{\bm{x}-\bm{x_l}} \leq \norm{\bm{x}-\bm{x_r}}\}$;~$S_r \leftarrow S \setminus S_l$\;
	$N.lc \leftarrow $~\textsf{BallTreeConstruct}($S_l, N_0$)\;
	$N.rc \leftarrow $~\textsf{BallTreeConstruct}($S_r, N_0$)\;
	\Return $N$\;
}
\end{algorithm}

\begin{algorithm}[t]
\caption{\textsf{Split}}
\label{alg:split}
\KwIn{subset $S \subset \mathcal{S}$;}
Select a random point $\bm{v} \in S$\;
$\bm{x_l} = \arg\max_{\bm{x} \in S} \norm{\bm{x}-\bm{v}}$;~$\bm{x_r} = \arg\max_{\bm{x} \in S} \norm{\bm{x}-\bm{x_l}}$\;
\Return $\bm{x_l}, \bm{x_r}$\;
\end{algorithm}

\paragraph{Ball-Tree Construction}
The Ball-Tree construction is shown in Algorithm \ref{alg:ball_construct}. 
We use the seed-grow rule (i.e., Algorithm \ref{alg:split}) to split a node $N$ into two: we randomly select a pair of pivot points $\bm{x_l},\bm{x_r} \in N.S$ that are furthest from each other; then, we partition each point $\bm{x} \in N.S$ to its closest pivot and split $S$ into two sets $S_l$ and $S_r$. 
Suppose $N_0$ is the maximum leaf size, and we use the data set $\mathcal{S}$ as input. The ball tree is built recursively until all leaf nodes have at most $N_0$ data points. 

\begin{theorem}[Construction Cost\cite{omohundro1989five}]
\label{theorem:ball_tree_construction}
The Ball-Tree can be constructed in $O(dn\log n)$ time and $O(nd)$ space.
\end{theorem}


According to Theorem \ref{theorem:ball_tree_construction}, the Ball-Tree construction is fast ($\tilde{O}(dn)$ time) and lightweight ($O(nd)$ space). In practice, the total number of nodes in Ball-Tree is usually less than $n$ as we often set $N_0$ much larger than 1. Thus, its space is usually less than $O(nd)$, which will be validated in Section \ref{sect:expt:index}.

\subsection{Ball-Tree for P2HNNS}
\label{sect:ball:search}
\vspace{-0.5em}
\paragraph{Node-Level Ball Bound} 
Before we present the search scheme of Ball-Tree to deal with the P2HNNS, we first develop a new lower bound for its nodes.

\begin{figure}[h]
\centering
\vspace{-0.75em}
\subfigure[Case 1]{
	\label{fig:node_lower_bound:case1}
	\includegraphics[width=0.21\textwidth]{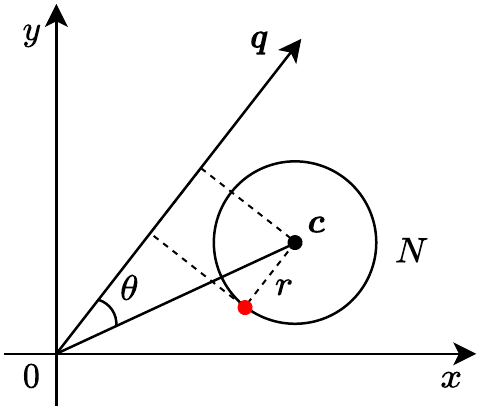}}%
\subfigure[Case 2]{
	\label{fig:node_lower_bound:case2}
	\includegraphics[width=0.21\textwidth]{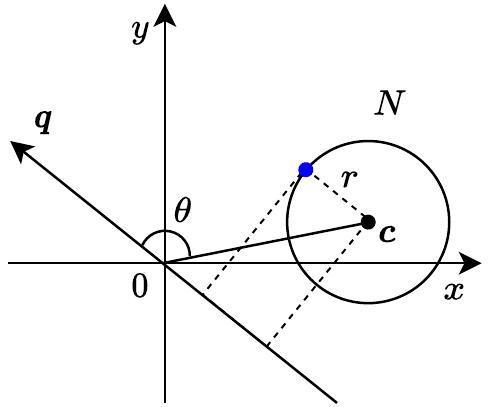}} 
\vspace{-0.5em}
\caption{An illustration of the node-level lower bound.}
\label{fig:node_lower_bound}
\vspace{-0.75em}
\end{figure}

\begin{theorem}[Node-Level Ball Bound]
\label{theorem:node_lower_bound}
Given a query $\bm{q}$ and a node $N$ that contains a set of data points $N.S$ centered at $N.\bm{c}$ with radius $N.r$, the minimum possible $\abso{\ip{\bm{x},\bm{q}}}$ of all data points $\bm{x} \in N.S$ and $\bm{q}$ is bounded as follows:
\begin{equation}
\label{eqn:ball_lower_bound}
\min_{\bm{x} \in N.S} \abso{\ip{\bm{x},\bm{q}}} \geq \max(\abso{\ip{\bm{q},N.\bm{c}}} - \norm{\bm{q}} \cdot N.r, 0).
\end{equation}
\end{theorem}
\begin{proof}
Let $\theta$ be the angle between $\bm{q}$ and $N.\bm{c}$. We have $\ip{\bm{q},N.\bm{c}} = \norm{\bm{q}} \norm{N.\bm{c}} \cos\theta$. Depending on the relationship of $\norm{N.\bm{c}} \cos\theta$ and $N.r$, the lower bound consists of three cases:
\begin{enumerate}[leftmargin=28pt,label*=(\arabic*)]
\item $\norm{N.\bm{c}} \cos\theta > N.r$. 
In this case, $\ip{\bm{x},\bm{q}} > 0$ for all $\bm{x} \in N.S$.
As shown in Figure \ref{fig:node_lower_bound:case1}, the red point that has the minimum projected distance $(\norm{N.\bm{c}} \cos\theta - N.r)$ to $\bm{q}$ is the one that contains the minimum $\abso{\ip{\bm{x},\bm{q}}}$, i.e., $\min_{\bm{x} \in N.S} \abso{\ip{\bm{x},\bm{q}}} \geq (\norm{N.\bm{c}} \cos\theta - N.r)\norm{\bm{q}} = \ip{\bm{q},N.\bm{c}}-\norm{\bm{q}} \cdot N.r$.

\item $\norm{N.\bm{c}} \cos\theta < -N.r$. In this case, $\ip{\bm{x},\bm{q}} < 0$ for all $\bm{x} \in N.S$. As shown in Figure \ref{fig:node_lower_bound:case2}, the blue point that has the minimum projected distance $(-\norm{N.\bm{c}} \cos\theta - N.r)$ to $\bm{q}$ is the one that contains the minimum $\abso{\ip{\bm{x},\bm{q}}}$, i.e., $\min_{\bm{x} \in N.S} \abso{\ip{\bm{x},\bm{q}}} \geq (-\norm{N.\bm{c}} \cos\theta - N.r)\norm{\bm{q}} = -\ip{\bm{q},N.\bm{c}}-\norm{\bm{q}} \cdot N.r$.

\item $\norm{N.\bm{c}} \cdot \abso{\cos\theta} \leq N.r$. In this case, as the virtual ball might contain some data points that are orthogonal to $\bm{q}$, the lower bound is 0.
\end{enumerate}

In summary, if $\norm{N.\bm{c}} \cdot \abso{\cos\theta} > N.r$, the lower bound is $\abso{\ip{\bm{q},N.\bm{c}}} - \norm{\bm{q}} \cdot N.r = \norm{\bm{q}}\norm{N.\bm{c}} \abso{\cos\theta} - \norm{\bm{q}}\cdot N.r = \norm{\bm{q}} (\norm{N.\bm{c}} \cdot \abso{\cos\theta} - N.r) > 0$; otherwise, the lower bound is 0. Thus, the lower bound is $\max(\abso{\ip{\bm{q},N.\bm{c}}} - \norm{\bm{q}} \cdot N.r, 0)$.
\end{proof}

\begin{algorithm}[t]
\caption{\textsf{BallTreeSearch}}
\label{alg:ball_search}
\KwIn{query $\bm{q}$, root node $N$;}
$\bm{q}.bm \leftarrow \emptyset$;~$\bm{q}.\lambda \leftarrow +\infty$\;
\textsf{SubBallTreeSearch}($\bm{q}$, $N$)\;
\Return $\bm{q}.bm$ and $\bm{q}.\lambda$\;
\SetKwFunction{Function}{\textsf{SubBallTreeSearch}}
\SetKwProg{Fn}{Function}{:}{}
\Fn{\Function{$\bm{q}$, $N$}}{
  $lb = \max(\abso{\ip{\bm{q}, N.\bm{c}}}-\norm{\bm{q}}\cdot N.r, 0)$\; \label{ball_tree_search：node_level_ball_bound}
  \If{$lb < \bm{q}.\lambda$}{ \label{ball_tree_search：sub:start}
	\If(\Comment*[f]{leaf node}){$\num{N} \leq N_0$}{
		\textsf{ExhaustiveScan}($\bm{q}, N$)\;
	} 
	\Else(\Comment*[f]{internal node}){
      Compute $\ip{\bm{q}, N.lc.\bm{c}}$ and $\ip{\bm{q}, N.rc.\bm{c}}$\; \label{branch:start}
	  \If {$\abso{\ip{\bm{q}, N.lc.\bm{c}}} < \abso{\ip{\bm{q}, N.rc.\bm{c}}}$} {
		\textsf{SubBallTreeSearch}($\bm{q}, N.lc$)\;
		\textsf{SubBallTreeSearch}($\bm{q}, N.rc$)\;
	  } \Else {
		\textsf{SubBallTreeSearch}($\bm{q}, N.rc$)\;
		\textsf{SubBallTreeSearch}($\bm{q}, N.lc$)\;
	  } \label{branch:end}
	} \label{ball_tree_search：sub:end}
  }
}
\SetKwFunction{Function}{\textsf{ExhaustiveScan}} 
\SetKwProg{Fn}{Function}{:}{}
\Fn{\Function{$\bm{q}$, $N$}}{ \label{exhau_scan:start}
  \ForEach{$\bm{x} \in N.S$} {
	\If {$\abso{\ip{\bm{x},\bm{q}}} < \bm{q}.\lambda$} {
      $\bm{q}.bm \leftarrow \bm{x}$;~$\bm{q}.\lambda \leftarrow \abso{\ip{\bm{x},\bm{q}}}$\;
    }
  } 
} \label{exhau_scan:end}
\end{algorithm}

\vspace{-0.6em}
\paragraph{Search Scheme}
We call the right-hand side (RHS) of Inequality \ref{eqn:ball_lower_bound} as the \emph{node-level ball bound}. 
With this lower bound, we can apply Ball-Tree to tackle the problem of P2HNNS by the branch-and-bound strategy. 
Suppose $\bm{q}.bm$ stores the best match data point (i.e., the closest point to the query $\bm{q}$) we found so far, and let $\bm{q}.\lambda$ be the current minimum $\abso{\ip{\bm{x},\bm{q}}}$. The search scheme is described in Algorithm \ref{alg:ball_search}. 

In Algorithm \ref{alg:ball_search}, we find the best match data point of $\bm{q}$ by traversing the tree in a depth-first manner. 
We first compute its node-level ball bound for each node $N$, i.e., $lb = \max(\abso{\ip{\bm{q}, N.\bm{c}}}-\norm{\bm{q}}\cdot N.r, 0)$ (Line \ref{ball_tree_search：node_level_ball_bound}). 
If this bound is at least the current minimum $\abso{\ip{\bm{x},\bm{q}}}$, i.e., $lb \geq \bm{q}.\lambda$, which means that this node cannot contain data points closer to $\bm{q}$, we prune this branch; otherwise, we continue to visit its branches (left child $N.lc$ and right child $N.lc$) based on some heuristic preferences (Lines \ref{ball_tree_search：sub:start}--\ref{ball_tree_search：sub:end}). 
If $N$ is a leaf node, we find the best match with an exhaustive scan (Lines \ref{exhau_scan:start}--\ref{exhau_scan:end}).

\paragraph{Branch Preference Choice}
To determine the branching order, we adopt the \emph{center preference}, which is based on the absolute inner products between $\bm{q}$ and the centers of $N.lc$ and $N.rc$ (Lines \ref{branch:start}--\ref{branch:end} in Algorithm \ref{alg:ball_search}) because we can roughly estimate the closeness between the data points and $\bm{q}$ by the closeness between the center and $\bm{q}$. 
Another choice is the \emph{lower bound preference}, which is based on the minimum possible node-level ball bounds for $N.lc$ and $N.rc$. 

If we only consider the two children of this node, the lower bound preference might be better than the center preference, as it can find the best match as soon as possible. 
However, if we consider traversing the tree, the lower bound preference might be easier to lead to the \emph{worse} branch at the early stage because the radii of the root and the first few nodes near the root  are usually very large, the node-level ball bounds for their children are probably all 0's. In this sense, the lower bound preference is worse than the center preference. 
We will further justify the two preference choices in Section \ref{sect:expt:preference}.

\paragraph{Limitations}
With the satisfied branch preference choice, Ball-Tree is efficient yet effective for P2HNNS. Nevertheless, there exist two limitations in Ball-Tree: 
\begin{enumerate}[leftmargin=28pt,label*=(\arabic*)]
\item We require an exhaustive scan through all data points in the leaf. Even though the points in each leaf are stored consecutively, which can be accessed sequentially, the total \emph{candidate verification cost} might be prohibitive as we might need to check many leaves. 

\item The centers in different internal and leaf nodes are not stored consecutively. When we compute the node-level ball bound between $\bm{q}$ and the center, we require once random access. Thus, a single \emph{node-level ball bound computation cost} is much more expensive than a single candidate verification cost.
\end{enumerate}

One can tune the leaf size $N_0$ to balance the total candidate verification cost and the total node-level ball bound computation cost, but it might not be helpful to reduce both costs. 
Next, we will propose a new tree structure, BC-Tree, that can reduce these two costs simultaneously. 

\section{BC-Tree}
\label{sect:bc}

\subsection{Overview}
\label{sect:bc:overview}
BC-Tree is built upon the Ball-Tree, which applies the same splitting rule to construct the tree recursively and maintains the same center and radius for their nodes. 
We add two virtual structures (i.e., ball and cone structures) for each data point in the leaf nodes of Ball-Tree. 
With the two structures, we can perform \emph{point-level pruning} to avoid the exhaustive scan in the leaf and reduce the total candidate verification cost. 

Moreover, with the linear properties of the center computation and the inner product computation, we design a new \emph{collaborative inner product computing} strategy. Using this strategy, we can cut down the total node-level ball bound computation cost by almost half. 
We will present the details of the two strategies in the following two subsections, respectively.

\subsection{Point-Level Pruning}
\label{sect:bc:leaf}
\begin{figure}[htb]
\centering
\vspace{-0.5em}
\includegraphics[width=0.24\textwidth]{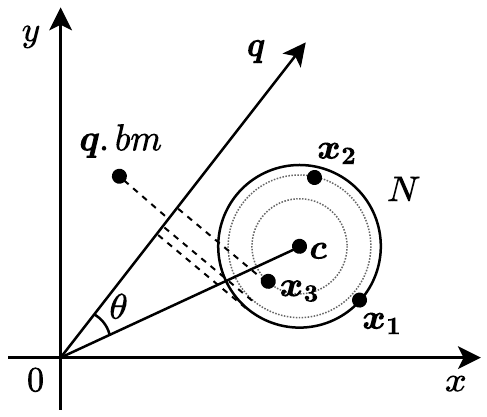}
\vspace{-0.75em}
\caption{An example of a leaf node $N$ with the ball structures of $\{\bm{x_1},\bm{x_2},\bm{x_3}\}$.}
\label{fig:point_lower_bound_ball}
\vspace{-0.25em}
\end{figure}

\vspace{-0.5em}
\paragraph{Point-Level Ball Bound}
To perform the point-level pruning, a natural idea is to apply the virtual ball structure for each data point in the leaf. As such, we can quickly get a lower bound for each data point. 
The advantage is that all virtual balls share the same center. Thus, given a leaf node $N$ with the center $N.\bm{c}$, we only need to maintain a radius $r_{\bm{x}}$ for each $\bm{x} \in N.S$, i.e., $r_{\bm{x}} = \norm{\bm{x} - N.\bm{c}}$. An example of the leaf node with the virtual ball structures is depicted in Figure \ref{fig:point_lower_bound_ball}. 

Based on the virtual ball structure, we now introduce a lower bound for each $\bm{x} \in N.S$ for point-level pruning. 
\begin{corollary}[Point-Level Ball Bound]
\label{corollary:point_lower_bound_ball}
Given a query $\bm{q}$ and a leaf node $N$ that maintains a center $N.\bm{c}$ and the radii $\{r_{\bm{x}}\}_{\bm{x} \in N.S}$, the minimum possible $\abso{\ip{\bm{x},\bm{q}}}$ of each data point $\bm{x} \in N.S$ and $\bm{q}$ is bounded as follows:
\begin{equation}
\label{eqn:ball_lower_bound_leaf}
\abso{\ip{\bm{x},\bm{q}}} \geq \max(\abso{\ip{\bm{q},N.\bm{c}}} - \norm{\bm{q}} \cdot r_{\bm{x}}, 0).
\end{equation}
\end{corollary}
\begin{proof}
The proof of Corollary \ref{corollary:point_lower_bound_ball} is similar to that of Theorem \ref{theorem:node_lower_bound}. To be concise, we omit the details here. 
\end{proof}

We call the RHS of Inequality \ref{eqn:ball_lower_bound_leaf} as the \emph{point-level ball bound}. 
Note that we have already computed $\ip{\bm{q}, N.\bm{c}}$ when we visit the leaf node $N$. Thus, the point-level ball bound for each $\bm{x} \in N.S$ can be computed in $O(1)$ time. 

Moreover, for the data points $\bm{x} \in N.S$, the terms $\ip{\bm{q},N.\bm{c}}$ and $\norm{\bm{q}}$ are constant for a certain $\bm{q}$. 
Thus, the point-level ball bound is a decreasing function of $r_{\bm{x}}$, i.e., this lower bound decreases as $r_{\bm{x}}$ increases. 
When we construct the BC-Tree, we sort the data points $\bm{x} \in N.S$ in descending order of $r_{\bm{x}}$. With this order, we can leverage this point-level ball bound to prune the data points \emph{in a batch manner}. 
For example, if this lower bound is at least $\bm{q}.\lambda$ for the current point, we do not need to verify this point and the remaining points as the lower bounds for the remaining points are also at least $\bm{q}.\lambda$. More details can be found in Sections \ref{sect:bc:construct} and \ref{sect:bc:search}.

This lower bound, however, only utilizes the center and radius of the ball structure but neglects the actual $l_2$ norm of the data point and its angle to $\bm{q}$. 
Next, we introduce a tighter lower bound for point-level pruning. 

\begin{figure}[h]
\centering
\vspace{-0.5em}
\includegraphics[width=0.24\textwidth]{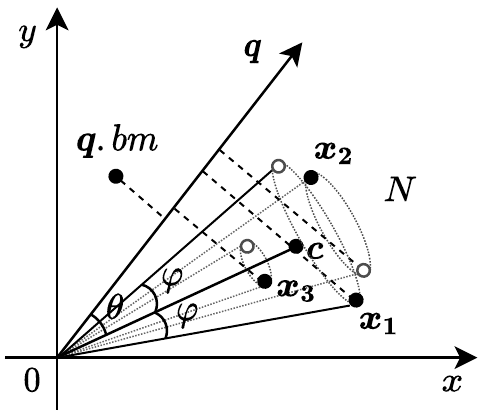}
\vspace{-0.75em}
\caption{An example of a leaf node $N$ with the cone structures of the same data points $\{\bm{x_1},\bm{x_2},\bm{x_3}\}$ as shown in Figure \ref{fig:point_lower_bound_ball}.}
\label{fig:point_lower_bound_cone}
\vspace{-0.25em}
\end{figure}

\paragraph{Point-Level Cone Bound}
To get a tighter bound, except for the virtual ball structure, we also maintain a virtual cone structure for each data point $\bm{x} \in N.S$, i.e., its $l_2$ norm $\norm{\bm{x}}$ and its angle $\varphi_{\bm{x}}$ to the center $N.\bm{c}$. An example of the leaf node with the virtual cone structures is shown in Figure \ref{fig:point_lower_bound_cone}. 

We continue to use $\theta$ to represent the angle between $N.\bm{c}$ and $\bm{q}$. 
Based on the virtual cone structure, we develop a new lower bound for each $\bm{x} \in N.S$ for point-level pruning. 
\begin{theorem}[Point-Level Cone Bound]
\label{theorem:point_lower_bound_cone}
Given a query $\bm{q}$ and a leaf node $N$ that maintains the $l_2$ norm $\norm{\bm{x}}$ and the angle $\varphi_{\bm{x}}$ to the center $N.\bm{c}$ for each $\bm{x} \in N.S$, the minimum possible $\abso{\ip{\bm{x},\bm{q}}}$ of each $\bm{x} \in N.S$ and $\bm{q}$ is bounded as below:
\begin{equation} \small
\label{eqn:cone_lower_bound}
\abso{\ip{\bm{x},\bm{q}}} \geq 
\begin{cases}
\norm{\bm{x}}\norm{\bm{q}}\cos(\theta+\varphi_{\bm{x}}),& \text{if }\cos(\theta+\varphi_{\bm{x}})>0  \\
& \text{ and } \cos\theta,\cos\varphi_{\bm{x}} > 0 \\
-\norm{\bm{x}}\norm{\bm{q}}\cos(\abso{\theta-\varphi_{\bm{x}}}),& \text{if }\cos(\abso{\theta-\varphi_{\bm{x}}})<0 \\
0, & \text{otherwise}
\end{cases}
\end{equation}
\end{theorem}
\begin{proof}
Let $\theta_{\bm{x},\bm{q}} \in [0,\pi]$ be the angle between $\bm{x}$ and $\bm{q}$. 
We have $\ip{\bm{x},\bm{q}} = \norm{\bm{x}} \norm{\bm{q}} \cos\theta_{\bm{x},\bm{q}}$. According to the triangle inequality, as $\varphi_{\bm{x}}$ is the angle between $\bm{x}$ and $N.\bm{c}$, we have the relationship $\abso{\theta - \varphi_{\bm{x}}} \leq \theta_{\bm{x},\bm{q}} \leq \theta + \varphi_{\bm{x}}$. As $0 \leq \theta,\varphi_{\bm{x}} \leq \pi$, we have $\sin \theta \geq 0$ and $\sin \varphi_{\bm{x}} \geq 0$. Thus,
\begin{align*}
\cos(\theta+\varphi_{\bm{x}})
& =    \cos\theta \cos\varphi_{\bm{x}} - \sin\theta \sin\varphi_{\bm{x}} \\
& \leq \cos\theta \cos\varphi_{\bm{x}} + \sin\theta \sin\varphi_{\bm{x}} \\
& =    \cos(\abso{\theta-\varphi_{\bm{x}}}).
\end{align*}

Since $\theta,\varphi_{\bm{x}} \in [0,\pi]$ and $\theta_{\bm{x},\bm{q}} \in [0,\pi]$, $(\theta + \varphi_{\bm{x}}) \in [\theta_{\bm{x},\bm{q}}, 2\pi]$.
\begin{enumerate}[leftmargin=28pt,label*=(\arabic*)]
\item We first consider $(\theta + \varphi_{\bm{x}}) \in [\theta_{\bm{x},\bm{q}}, \pi]$. Since $0 \leq \abso{\theta - \varphi_{\bm{x}}} \leq \theta_{\bm{x},\bm{q}} \leq \theta + \varphi_{\bm{x}} \leq \pi$, $\cos\theta_{\bm{x},\bm{q}}$ decreases monotonically as $\theta_{\bm{x},\bm{q}}$ increases. Thus, we have 
$\cos(\theta+\varphi_{\bm{x}}) \leq \cos\theta_{\bm{x},\bm{q}} \leq \cos(\abso{\theta-\varphi_{\bm{x}}})$. 
The lower bound consists of three cases: 
\begin{enumerate}[leftmargin=14pt] 
\item $\cos(\theta+\varphi_{\bm{x}})>0$. Since $\cos(\abso{\theta-\varphi_{\bm{x}}}) \geq \cos\theta_{\bm{x},\bm{q}} \geq \cos(\theta + \varphi_{\bm{x}}) > 0$, the lower bound of $\abso{\ip{\bm{x},\bm{q}}}$ is $\norm{\bm{x}} \norm{\bm{q}} \cos(\theta + \varphi_{\bm{x}})$.

\item $\cos(\abso{\theta-\varphi_{\bm{x}}})<0$. As $\cos(\theta + \varphi_{\bm{x}}) \leq \cos\theta_{\bm{x},\bm{q}} \leq \cos(\abso{\theta - \varphi_{\bm{x}}}) < 0$, the lower bound is $-\norm{\bm{x}}\norm{\bm{q}} \cos(\abso{\theta - \varphi_{\bm{x}}})$.

\item $\cos(\theta+\varphi_{\bm{x}}) \leq 0$ and $\cos(\abso{\theta-\varphi_{\bm{x}}}) \geq 0$. As $\bm{x}$ might be orthogonal to $\bm{q}$, the lower bound is $0$.
\end{enumerate}

For case a), since $\cos(\theta+\varphi_{\bm{x}})>0$ and we consider $(\theta + \varphi_{\bm{x}}) < \pi$, $(\theta + \varphi_{\bm{x}})$ is smaller than $\tfrac{\pi}{2}$. Thus, both $\cos\theta>0$ and $\cos\varphi_{\bm{x}} > 0$ are valid for this case. 

\item We then consider $(\theta+\varphi_{\bm{x}}) \in (\pi, 2\pi]$. For such case, the range of $\theta_{\bm{x},\bm{q}}$ is $\abso{\theta - \varphi_{\bm{x}}} \leq \theta_{\bm{x},\bm{q}} \leq \pi$. Thus, we have $-1 \leq \cos\theta_{\bm{x},\bm{q}} \leq \cos(\abso{\theta-\varphi_{\bm{x}}})$. 
The lower bound consists of two cases:
\begin{enumerate}[leftmargin=14pt] 
\item $\cos(\abso{\theta-\varphi_{\bm{x}}})<0$. 
As $-1 \leq \cos\theta_{\bm{x},\bm{q}} \leq \cos(\abso{\theta - \varphi_{\bm{x}}}) < 0$, the lower bound of $\abso{\ip{\bm{x},\bm{q}}}$ is $-\norm{\bm{x}}\norm{\bm{q}} \cos(\abso{\theta - \varphi_{\bm{x}}})$.

\item $\cos(\abso{\theta-\varphi_{\bm{x}}}) \geq 0$. 
As $\bm{x}$ might be orthogonal to $\bm{q}$, the lower bound is $0$.
\end{enumerate}

Note that as $(\theta+\varphi_{\bm{x}}) \in (\pi, 2\pi]$, the conditions $\cos\theta > 0$ and $\cos \varphi_{\bm{x}} > 0$ cannot be valid simultaneously. 
\end{enumerate}

In summary, if $\cos(\theta+\varphi_{\bm{x}})>0$ and $\cos\theta > 0$ and $\cos\varphi_{\bm{x}} > 0$, the lower bound is $\norm{\bm{x}} \norm{\bm{q}} \cos(\theta + \varphi_{\bm{x}})$; otherwise, if $\cos(\abso{\theta-\varphi_{\bm{x}}})<0$, the lower bound is $-\norm{\bm{x}}\norm{\bm{q}} \cos(\abso{\theta - \varphi_{\bm{x}}})$; otherwise, the lower bound is 0. Theorem \ref{theorem:point_lower_bound_cone} is proved.
\end{proof}

We call the RHS of Inequality \ref{eqn:cone_lower_bound} as the \emph{point-level cone bound}. 
We can infer that $\norm{\bm{q}}\norm{\bm{x}}\cos(\theta+\varphi_{\bm{x}}) = \norm{\bm{q}}\cos\theta \cdot \norm{\bm{x}}\cos\varphi_{\bm{x}} - \norm{\bm{q}}\sin\theta \cdot \norm{\bm{x}}\sin\varphi_{\bm{x}}$ and $
\norm{\bm{q}}\norm{\bm{x}}\cos(\abso{\theta-\varphi_{\bm{x}}}) = \norm{\bm{q}}\cos\theta \cdot \norm{\bm{x}}\cos\varphi_{\bm{x}} + \norm{\bm{q}}\sin\theta \cdot \norm{\bm{x}}\sin\varphi_{\bm{x}}$. 
When we construct the BC-Tree, we compute $\norm{\bm{x}}$ and $\varphi_{\bm{x}}$ and store $\norm{\bm{x}}\cos\varphi_{\bm{x}}$ and $\norm{\bm{x}}\sin\varphi_{\bm{x}}$ for each $\bm{x} \in N.S$. 
Moreover, as we have already computed $\ip{\bm{q},N.\bm{c}}$ when we visit the leaf node, we can compute $\norm{\bm{q}}\cos\theta = \ip{\bm{q},N.\bm{c}}/\norm{N.\bm{c}}$ and $\norm{\bm{q}}\sin\theta = \sqrt{\norm{\bm{q}}^2 - \norm{\bm{q}}^2\cos^2 \theta}$ in $O(1)$ time. Thus, this point-level cone bound can be computed in $O(1)$ time. More details can be found in Sections \ref{sect:bc:construct} and \ref{sect:bc:search}.

We now demonstrate that the point-level cone bound is tighter than the point-level ball bound. 
\begin{theorem}
\label{theorem:tight_lower_bound}
Given a query $\bm{q}$, for any data point $\bm{x}$ in a leaf node $N$, its point-level cone bound is tighter than the point-level ball bound.
\end{theorem}
\begin{proof}[Proof Sketch]
We continue to use the notations in Corollary \ref{corollary:point_lower_bound_ball} and
Theorem \ref{theorem:point_lower_bound_cone}. To prove this theorem, we should demonstrate that the RHS of Inequality \ref{eqn:cone_lower_bound} is at least the RHS of Inequality \ref{eqn:ball_lower_bound_leaf}. 
It should be noted that both bounds contain three cases; there might be nine cases in total. Nevertheless, some cases do not happen because their conditions conflict. 

We now show that 
$\norm{\bm{x}} \norm{\bm{q}} \cos(\theta+\varphi_{\bm{x}}) \geq \ip{\bm{q},N.\bm{c}} - \norm{\bm{q}} \cdot r_{\bm{x}}$ under the conditions $\cos(\theta+\varphi_{\bm{x}})>0$, $\cos\theta>0$, $\cos \varphi_{\bm{x}}>0$, and $\norm{N.\bm{c}} \cos\theta > r_{\bm{x}}$. After removing $\norm{\bm{q}}$ on both sides, we show that $\norm{\bm{x}} \cos(\theta+\varphi_{\bm{x}}) \geq \norm{N.\bm{c}}\cos\theta - r_{\bm{x}}$ is valid because
\begin{displaymath} 
\begin{split} 
   & \norm{\bm{x}} \cos(\theta+\varphi_{\bm{x}}) - \norm{N.\bm{c}}\cos\theta + r_{\bm{x}} \\
= ~& r_{\bm{x}} - \norm{N.\bm{c}}\cos\theta + \norm{\bm{x}} (\cos\theta \cos\varphi_{\bm{x}} - \sin\theta \sin\varphi_{\bm{x}}) \\
= ~& r_{\bm{x}} - (\norm{\bm{x}}\sin\varphi_{\bm{x}} \cdot \sin\theta + (\norm{N.\bm{c}} - \norm{\bm{x}} \cos\varphi_{\bm{x}}) \cdot \cos\theta) \\
\geq ~& r_{\bm{x}} - \sqrt{(\norm{\bm{x}} \sin\varphi_{\bm{x}})^2 + (\norm{N.\bm{c}} - \norm{\bm{x}} \cos\varphi_{\bm{x}})^2} \cdot \sqrt{1} \\ 
= ~& r_{\bm{x}} - r_{\bm{x}} = 0
\end{split}
\end{displaymath}

The last second step is based on the Cauchy–Schwarz Inequality. The last step is based on Pythagoras' Theorem, and the illustration is depicted in Figure \ref{fig:point_lower_bound_cone}. 

\begin{figure}[t]
\centering
\includegraphics[width=0.24\textwidth]{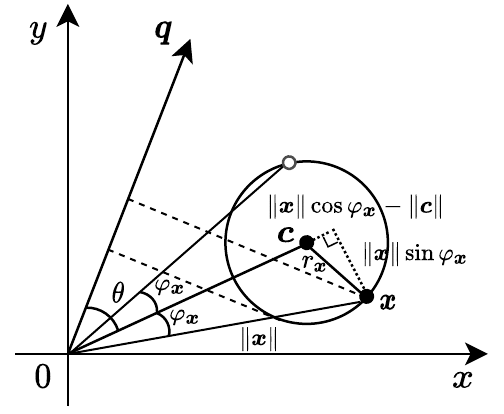}
\vspace{-0.75em}
\caption{An illustration of the point-level ball bound and point-level cone bound of a data point $\bm{x}$ in the leaf node. We observe that $(\norm{\bm{x}} \sin\varphi_{\bm{x}})^2 + (\norm{N.\bm{c}} - \norm{\bm{x}} \cos\varphi_{\bm{x}})^2 = r_{\bm{x}}^2$.}
\label{fig:point_ball_cone}
\vspace{-0.75em}
\end{figure}

The proofs for the rest cases are similar to that of this case. To be concise, we omit the details here.
\end{proof}

Theorem \ref{theorem:tight_lower_bound} is verified by Figures \ref{fig:point_lower_bound_ball} and \ref{fig:point_lower_bound_cone}. For the leaf node $N$ with the same $\{\bm{x_1},\bm{x_2},\bm{x_3}\}$ and $\bm{q}$, only $\bm{x_3}$ is pruned by its point-level ball bound, while all points are pruned by their point-level cone bounds. We will validate the effectiveness of the two point-level lower bounds in Section \ref{sect:expt:lower_bound}.

Until now, we have presented two new lower bounds for point-level pruning in $O(1)$ time, which can be utilized to reduce the total candidate verification cost. Next, we introduce the collaborative inner product computing strategy, which aims to reduce the node-level ball bound computation cost. 

\subsection{Collaborative Inner Product Computing}
\label{sect:bc:cipc}
The collaborative inner product computing strategy utilizes the linear properties of the center computation and inner product computation. According to Equation \ref{eqn:center}, we first present the linear property of the center computation:
\begin{lemma}
\label{lemma:center_linear_property}
Given an arbitrary internal node $N$ and its two children $N.lc$ and $N.rc$, we have
\begin{equation}
\label{eqn:center-linear}
N.\bm{c} \cdot \num{N} = N.lc.\bm{c} \cdot \num{N.lc} + N.rc.\bm{c} \cdot \num{N.rc}.
\end{equation}
\end{lemma}

According to Lemma \ref{lemma:center_linear_property}, given an internal node $N$, once the centers of its two children are determined, its center $N.\bm{c}$ can be computed in $O(d)$ time, which can be used to speed up the BC-Tree construction. More details will be presented in Section \ref{sect:bc:construct}. In this subsection, we use its another expression:
\begin{equation}
\label{eqn:center-linear-2}
N.rc.\bm{c} = \tfrac{\num{N}}{\num{N.rc}} \cdot N.\bm{c} - \tfrac{\num{N.lc}}{\num{N.rc}} \cdot N.lc.\bm{c}.
\end{equation}

Based on Equation \ref{eqn:center-linear-2}, we present the linear property of the inner product computation for a node and its two children.
\begin{lemma}
\label{lemma:ip_linear_property}
Given a query $\bm{q}$, an internal node $N$ and its two children $N.lc$ and $N.rc$, the inner products of $\bm{q}$ and the three nodes $N$, $N.lc$, and $N.rc$ satisfy the following relationship:
\begin{equation}
\label{eqn:ip_reduce}
\ip{\bm{q}, N.rc.\bm{c}} = \tfrac{\num{N}}{\num{N.rc}} \cdot \ip{\bm{q}, N.\bm{c}} - \tfrac{\num{N.lc}}{\num{N.rc}} \cdot \ip{\bm{q}, N.lc.\bm{c}}.
\end{equation}
\end{lemma}

Given a query $\bm{q}$, when we visit an internal node $N$, we have computed $\ip{\bm{q},N.\bm{c}}$. 
If the node-level ball bound fails, we need to compute the inner products of $\bm{q}$ and the centers of its two children $N.lc$ and $N.rc$ (Line \ref{branch:start} in Algorithm \ref{alg:ball_search}) to determine the preference. 
Suppose we have computed $\ip{\bm{q},N.lc.\bm{c}}$ for its left child $N.lc$. According to Lemma \ref{lemma:ip_linear_property}, the $\ip{\bm{q},N.rc.\bm{c}}$ for its right child $N.rc$ can be computed in $O(1)$ time. 
We call this strategy \emph{collaborative inner product computing}.

Note that the direct cost of the node-level ball bound is the inner product computation of the query and the center. 
With this strategy, the node-level ball bound for $N.rc$ can also be computed in $O(1)$ time. 
Suppose $C_{N}$ is the total number of inner product computations for this bound when we traverse the tree. We show that $C_{N}$ can be reduced by almost half. 
\begin{theorem}
\label{theorem:lower_bound_cost_reduce}
$C_{N}$ can be reduced to $\tfrac{C_{N}+1}{2}$ with Lemma \ref{lemma:ip_linear_property}.
\end{theorem}
\begin{proof}
$C_{N}$ is an odd number because we must compute the node-level ball bound (with once inner product computation) for the root. When we traverse the tree for the remaining nodes, we require twice inner product computations for its left child and right child or prune it directly. With Lemma \ref{lemma:ip_linear_property}, we only need to compute once inner product for the left child. Thus, $C_{N}$ can be reduced to $\tfrac{C_{N}+1}{2}$. Theorem \ref{theorem:lower_bound_cost_reduce} is proved.
\end{proof}
\vspace{-0.5em}

\begin{algorithm}[t]
\caption{\textsf{BCTreeConstruct}}
\label{alg:bc_construct}
\KwIn{subset $S \subset \mathcal{S}$, maximum leaf size $N_0$;}
$N.S \leftarrow S$\;
\If(\Comment*[f]{leaf node}){$\num{N} \leq N_0$} {
	$N.\bm{c} = \frac{1}{\num{N}} \sum_{\bm{x} \in N.S} \bm{x}$\; \label{bc:leaf:center_radius:start}
	$N.r = \max_{\bm{x} \in N.S} \norm{\bm{x}-N.\bm{c}}$\; \label{bc:leaf:center_radius:end}
	\ForEach{$\bm{x} \in N.S$} {
		Compute and store $\norm{\bm{x}}$ and $r_{\bm{x}}$\; \label{bc:leaf:r_x}
		$\cos\varphi_{\bm{x}} = \tfrac{\ip{\bm{x}, N.\bm{c}}}{\norm{\bm{x}} \cdot \norm{N.\bm{c}}}$; $\sin\varphi_{\bm{x}} = \sqrt{1 - \cos^2 \varphi_{\bm{x}}}$\; \label{bc:leaf:cone:start}
		Compute and store $\norm{\bm{x}}\cos\varphi_{\bm{x}}$ and $\norm{\bm{x}}\sin\varphi_{\bm{x}}$\; \label{bc:leaf:cone:end}
	}
	Sort $\bm{x} \in N.S$ in descending order of $r_{\bm{x}}$\;  \label{bc:leaf:sort_r_x}
	\Return $N$\;
}
\Else(\Comment*[f]{internal node}){
	$\bm{x_l}, \bm{x_r} \leftarrow $~\textsf{Split}($S$)\;
	$S_l \leftarrow \{\bm{x} \in S \mid \norm{\bm{x}-\bm{x_l}} \leq \norm{\bm{x}-\bm{x_r}}\}$;~$S_r \leftarrow S \setminus S_l$\;
	$N.lc \leftarrow $~\textsf{BCTreeConstruct}($S_l, N_0$)\;
	$N.rc \leftarrow $~\textsf{BCTreeConstruct}($S_r, N_0$)\;
	Compute and store $N.\bm{c}$ based on Lemma \ref{lemma:center_linear_property}\; \label{bc:internal:center_radius:start}
	$N.r = \max_{\bm{x} \in S} \norm{\bm{x}-N.\bm{c}}$\; \label{bc:internal:center_radius:end}
	\Return $N$\;
}
\end{algorithm}

\subsection{BC-Tree Construction}
\label{sect:bc:construct}
The BC-Tree construction is depicted in Algorithm \ref{alg:bc_construct}. Like Algorithm \ref{alg:ball_construct}, we use Algorithm \ref{alg:split} for splitting, and we maintain the center and radius for the internal and leaf nodes of BC-Tree. The differences  in Algorithm \ref{alg:bc_construct} are listed as below:
\begin{itemize}
\item We separate the computation of center and radius for leaf nodes (Lines \ref{bc:leaf:center_radius:start}--\ref{bc:leaf:center_radius:end}) and internal nodes (Lines \ref{bc:internal:center_radius:start}--\ref{bc:internal:center_radius:end}). Based on Lemma \ref{lemma:center_linear_property}, the center computation time for each internal node $N$ can be reduced from $O(d\num{N})$ to $O(d)$.

\item We compute $r_{\bm{x}}$ and $\norm{\bm{x}}$ for each $\bm{x} \in N.S$ (Line \ref{bc:leaf:r_x}) and sort all $\bm{x} \in N.S$ in descending order of $r_{\bm{x}}$ (Line \ref{bc:leaf:sort_r_x}) to utilize point-level ball bound for pruning.

\item To use point-level cone bound, we determine $\norm{\bm{x}} \cos\varphi_{\bm{x}}$ and $\norm{\bm{x}} \cos\varphi_{\bm{x}}$ for each $\bm{x} \in N.S$ (Lines \ref{bc:leaf:cone:start}--\ref{bc:leaf:cone:end}).
\end{itemize}

\vspace{-0.8em}
\begin{theorem}
\label{theorem:bc_tree_construction}
The BC-Tree can be constructed in $O(dn\log n)$ time and $O(nd)$ space.
\end{theorem}
\begin{proof}
\vspace{-0.2em}
For the internal node $N$ of BC-Tree, we can reduce the time to compute the center, but its time complexity is still $O(d\num{N})$ as it uses the same splitting rule as Ball-Tree. 
For the leaf node $N$, besides the center and radius, it maintains $r_{\bm{x}}$, $\norm{\bm{x}} \cos\varphi_{\bm{x}}$, and $\norm{\bm{x}} \cos\varphi_{\bm{x}}$ for each $\bm{x} \in N.S$. Such computations take $O(d\num{N})$ time. Thus, like Ball-Tree, the time to construct a node of BC-Tree is $O(d\num{N})$, and the time to construct the whole BC-Tree is $O(dn\log n)$. 

For the memory usage, except for $O(nd)$ space to store the centers of all nodes, BC-Tree uses $3$ $n$-size arrays to store $r_{\bm{x}}$, $\norm{\bm{x}} \cos\varphi_{\bm{x}}$, and $\norm{\bm{x}} \cos\varphi_{\bm{x}}$ for all $\bm{x} \in \mathcal{S}$. Thus, the total space of BC-Tree is $O(nd+3n) = O(nd)$. 
\end{proof}
\vspace{-0.5em}

According to Theorem \ref{theorem:bc_tree_construction}, BC-Tree is constructed as efficiently and lightweight as Ball-Tree. In practice, BC-Tree can be constructed faster  than Ball-Tree, but it uses a larger memory cost. We will validate this analysis in Section \ref{sect:expt:index}. 

\begin{algorithm}[t]
\caption{\textsf{BCTreeSearch}}
\label{alg:bc_search}
\KwIn{query $\bm{q}$, root node $N$;}
$\bm{q}.bm \leftarrow \emptyset$;~$\bm{q}.\lambda \leftarrow +\infty$;~$ip_{node} \leftarrow \ip{\bm{q}, N.\bm{c}}$\;
\textsf{SubBCTreeSearch}($\bm{q}, N, ip_{node}$)\;
\Return $\bm{q}.bm$ and $\bm{q}.\lambda$\;
\SetKwFunction{Function}{\textsf{SubBCTreeSearch}}
\SetKwProg{Fn}{Function}{:}{}
\Fn{\Function{$\bm{q}$, $N$, $ip_{node}$}}{
  $lb \leftarrow \max(\abso{ip_{node}}-\norm{\bm{q}}\cdot N.r, 0)$\;\label{recursive_bc_search:lb}
  \If{$lb < \bm{q}.\lambda$}{
    \If(\Comment*[f]{leaf node}){$\num{N} \leq N_0$} {
      \textsf{ScanWithPruning}($\bm{q}, N, ip_{node}$)\;
    } 
    \Else(\Comment*[f]{internal node}){
      $ip_{lc} \gets \ip{\bm{q}, N.lc.\bm{c}}$\; \label{recursive_bc_search:branch_order:start}
      $ip_{rc} \gets \ip{\bm{q}, N.rc.\bm{c}}$ by Equation \ref{eqn:ip_reduce}\; \label{recursive_bc_search:branch_order:end}
      \If {$\abso{ip_{lc}} < \abso{ip_{rc}}$} {
        \textsf{SubBCTreeSearch}($\bm{q}, N.lc, ip_{lc}$)\;
		\textsf{SubBCTreeSearch}($\bm{q}, N.rc, ip_{rc}$)\;
	  } \Else {
		\textsf{SubBCTreeSearch}($\bm{q}, N.rc, ip_{rc}$)\;
		\textsf{SubBCTreeSearch}($\bm{q}, N.lc, ip_{lc}$)\;
	  }
	}
  }
}
\SetKwFunction{Function}{\textsf{ScanWithPruning}}
\SetKwProg{Fn}{Function}{:}{}
\Fn{\Function{$\bm{q}$, $N$, $ip_{node}$}}{
  $\cos\theta = ip_{node}/\norm{N.\bm{c}}$; $\sin\theta = \sqrt{1 - \cos^2 \theta}$\;
  \ForEach{$\bm{x} \in N.S$} {
	$lb_{ball} = \max(\abso{ip_{node}}-\norm{\bm{q}}\cdot N.r, 0)$\; \label{bc_linear:point_level_ball_bound:start}
	\lIf{$lb_{ball} \geq \bm{q}.\lambda$}{\textbf{break}} \label{bc_linear:point_level_ball_bound:end}
	Compute $lb_{cone}$ by the RHS of Inequality \ref{eqn:cone_lower_bound}\; \label{bc_linear:point_level_cone_bound:start}
	\If {$lb_{cone} < \bm{q}.\lambda$} {
	  \If {$\abso{\ip{\bm{q},\bm{x}}} < \bm{q}.\lambda$} {
	    $\bm{q}.bm \leftarrow \bm{x}$;~$\bm{q}.\lambda \leftarrow \abso{\ip{\bm{q},\bm{x}}}$\;
	  }
	} \label{bc_linear:point_level_cone_bound:end}
  }
}
\end{algorithm}

\subsection{BC-Tree For P2HNNS}
\label{sect:bc:search}
The search scheme of BC-Tree is presented in Algorithm \ref{alg:bc_search}, where the differences from Ball-Tree are described as below.
\begin{itemize}
\item We apply the collaborative inner product computing strategy to reduce the total node-level ball bound computation cost. 
The input $ip_{node}$ is the inner product of $\bm{q}$ and the center $N.\bm{c}$ in this node. 
As $ip_{node}$ has been computed, the node-level ball bound $lb$ can be computed in $O(1)$ time (Line \ref{recursive_bc_search:lb}). 
To determine the branch order, we first take $O(d)$ time to compute $\ip{\bm{q}, N.lc.\bm{c}}$; then according to Lemma \ref{lemma:ip_linear_property}, $\ip{\bm{q}, N.rc.\bm{c}}$ can be determined by $ip_{node}$ and $\ip{\bm{q}, N.lc.\bm{c}}$ in $O(1)$ time (Lines \ref{recursive_bc_search:branch_order:start}--\ref{recursive_bc_search:branch_order:end}). 

\item We perform the point-level pruning to reduce the total candidate verification cost. 
We first apply the point-level ball bound $lb_{ball}$, as it can prune data points in a batch (Lines \ref{bc_linear:point_level_ball_bound:start}--\ref{bc_linear:point_level_ball_bound:end}). 
If it fails, we apply the tighter point-level cone bound $lb_{cone}$ for pruning (Lines \ref{bc_linear:point_level_cone_bound:start}--\ref{bc_linear:point_level_cone_bound:end}). 
Note that as $lb_{cone}$ is not an increasing (or decreasing) function of $\norm{\bm{x}}$ (or $\varphi_{\bm{x}}$), it is only effective for a single point $\bm{x}$. 
\end{itemize}

\section{Experiments}
\label{sect:expt}
In this section, we study the performance of Ball-Tree and BC-Tree for solving P2HNNS. We focus on the in-memory workload. 
All methods are written in C++ and compiled by g++-8 using -O3 optimization. We conduct all experiments in a single thread on a machine with Intel\textsuperscript{\textregistered} Xeon\textsuperscript{\textregistered} Platinum 8170 CPU@2.10GHz and 512 GB memory, running on CentOS 7.4.

\begin{table}[t]
\renewcommand{\arraystretch}{1.0}
\centering
\caption{Statistics of data sets.}
\label{table:dataset}
\vspace{-0.5em}
\begin{tabular}{>{\hfil}p{40pt}<{\hfil} >{\hfil}p{45pt}<{\hfil} c >{\hfil}p{35pt}<{\hfil}c} \toprule
Data Sets & $n$ & $d$ & Data Size & Data Type \\ \midrule
Music    & 1,000,000   & 100   & 386 MB & Rating  \\
GloVe    & 1,183,514   & 100   & 460 MB & Text    \\ 
Sift     & 985,462     & 128   & 485 MB & Image   \\
UKBench  & 1,097,907   & 128   & 541 MB & Image   \\
Tiny     & 1,000,000   & 384   & 1.5 GB & Image   \\
Msong    & 992,272     & 420   & 1.6 GB & Audio   \\ 
NUSW     & 268,643     & 500   & 514 MB & Image   \\
Cifar-10 & 50,000      & 512   & 98  MB & Image   \\
Sun      & 79,106      & 512   & 155 MB & Image   \\
LabelMe  & 181,093     & 512   & 355 MB & Image   \\
Gist     & 982,694     & 960   & 3.6 GB & Image   \\
Enron    & 94,987      & 1,369 & 497 MB & Text    \\ 
Trevi    & 100,900     & 4,096 & 1.6 GB & Image   \\
P53      & 31,153      & 5,408 & 643 MB & Biology \\ \cmidrule(lr){1-5}
Deep100M & 100,000,000 & 96    & 36.1 GB & Image  \\
Sift100M & 99,986,452  & 128   & 48.0 GB & Image  \\
\bottomrule
\end{tabular}
\vspace{-1.0em}
\end{table}

\begin{table*}
\renewcommand{\arraystretch}{1.0}
\caption{The indexing time (Time, in Seconds) and index size (Size, in Megabytes) of Ball-Tree, BC-Tree, FH, and NH.}
\vspace{-0.5em}
\label{tab:index}
\centering
\begin{tabular}{ccccccccccccc} \toprule
\multirow{2}{*}{Data Sets} & 
\multicolumn{2}{c}{BC-Tree} & 
\multicolumn{2}{c}{Ball-Tree} & 
\multicolumn{2}{c}{NH ($\lambda=d$)} & 
\multicolumn{2}{c}{NH ($\lambda=8d$)} & 
\multicolumn{2}{c}{FH ($\lambda=d$)} & 
\multicolumn{2}{c}{FH ($\lambda=8d$)} \\ 
\cmidrule(lr){2-3} \cmidrule(lr){4-5} \cmidrule(lr){6-7} 
\cmidrule(lr){8-9} \cmidrule(lr){10-11} \cmidrule(lr){12-13}
& Time & Size & Time & Size & Time & Size & Time & Size & Time & Size & Time & Size \\ \midrule
Music    & \textbf{10.5}  & 35.4  & 12.6  & \textbf{23.0}  & 117.3 & 1,471.2 & 194.2   & 1,471.2 & 32.1    & 1,096.0  & 106.9   & 1,096.0  \\ 
GloVe    & \textbf{15.3}  & 40.4  & 18.4  & \textbf{25.7}  & 131.4 & 1,753.9 & 202.7   & 1,753.9 & 37.3    & 1,307.4  & 115.7   & 1,309.8  \\ 
Sift     & \textbf{12.0}  & 34.0  & 13.0  & \textbf{21.9}  & 124.9 & 1,451.4 & 283.8   & 1,451.4 & 54.3    & 1,134.0  & 222.8   & 1,138.1  \\ 
UKBench  & \textbf{12.7}  & 38.7  & 13.4  & \textbf{25.2}  & 137.2 & 1,616.6 & 329.2   & 1,616.6 & 59.4    & 1,264.7  & 251.8   & 1,268.8  \\
Tiny     & \textbf{37.8}  & 69.5  & 42.5  & \textbf{57.3}  & 266.4 & 1,504.9 & 1,080.5 & 1,504.9 & 245.1   & 2,504.5  & 953.8   & 2,649.6  \\ 
Msong    & \textbf{77.2}  & 82.3  & 83.3  & \textbf{70.0}  & 157.6 & 1,500.7 & 475.3   & 1,500.7 & 106.2   & 3,011.6  & 427.9   & 3,141.7  \\ 
NUSW     & \textbf{26.0}  & 26.8  & 30.3  & \textbf{23.4}  & 42.9  & 456.0   & 132.8   & 456.0   & 34.3    & 1,123.3  & 114.2   & 1,184.5  \\ 
Cifar-10 & \textbf{1.6}   & 4.2   & \textbf{1.6} & \textbf{3.6} & 16.7 & 137.8 & 59.6  & 137.8   & 16.0    & 500.1    & 52.9    & 757.6    \\ 
Sun      & \textbf{3.0}   & 7.1   & 3.1   & \textbf{6.1}   & 36.1  & 180.6   & 189.5   & 180.6   & 31.9    & 528.7    & 153.3   & 850.5    \\ 
LabelMe  & \textbf{8.3}   & 16.9  & 8.6   & \textbf{14.7}  & 71.3  & 330.3   & 418.9   & 330.3   & 73.2    & 757.3    & 355.6   & 1,014.8  \\ 
Gist     & \textbf{111.4} & 150.4 & 122.3 & \textbf{138.3} & 817.5 & 1,669.0 & 5,157.8 & 1,669.0 & 1,002.3 & 9,993.3  & 6,469.9 & 11,121.8 \\ 
Enron    & \textbf{27.8}  & 37.8  & 29.6  & \textbf{36.6}  & 41.2  & 598.1   & 143.4   & 598.1   & 67.6    & 2,389.4  & 191.0   & 2,847.9  \\
Trevi    & \textbf{28.0}  & 62.1  & 28.9  & \textbf{60.9}  & 415.7 & 4,247.2 & 1,477.5 & 4,247.2 & 1,630.9 & 61,615.8 & 2,533.1 & 73,912.9 \\ 
P53      & \textbf{10.9}  & 29.4  & 11.4  & \textbf{29.0}  & 313.6 & 7,190.0 & 1,023.2 & 7,190.0 & 1,127.4 & 57,239.9 & 1988.1  & 71,528.4 \\ \cmidrule(lr){1-13}
Deep100M & \textbf{2,813.9} & 3,116.3 & 3,167.1 & \textbf{1,890.4} & 19,138.2 & 146,868.1 & 28,766.0 & 146,868.1 & 3,811.6 & 98,269.9 & 14,655.0 & 98,269.9 \\ 
Sift100M & \textbf{3,060.0} & 3,422.3 & 3,225.6 & \textbf{2,201.7} & 22,420.6 & 146,850.0 & 39,870.7 & 146,850.0 & 5,316.3 & 98,433.9 & 22,164.4 & 98,433.9 \\
\bottomrule
\end{tabular}
\vspace{-1.0em}
\end{table*}

\vspace{-0.1em}
\subsection{Data Sets and Queries}
\label{sect:expt:datasets}
\vspace{-0.1em}
In the experiments, we choose fourteen real-world data sets, i.e., 
Music \cite{morozov2018non}, 
GloVe,\footnote{\url{https://nlp.stanford.edu/projects/glove/}.} 
Sift,\footnote{\url{http://corpus-texmex.irisa.fr/}.} 
UKBench \cite{nister2006scalable}, 
Tiny \cite{torralba200880}, 
Msong,\footnote{\url{http://www.ifs.tuwien.ac.at/mir/msd/download.html}.} 
NUSW \cite{chua2009nus}, 
Cifar-10 \cite{krizhevsky2009learning},  
Sun,\footnote{\url{https://github.com/DBAIWangGroup/nns_benchmark/tree/master/data}.}
LabelMe \cite{russell2008labelme}, 
Gist,\footnote{\url{http://corpus-texmex.irisa.fr/}.} 
Enron,\footnote{\url{https://www.cs.cmu.edu/~enron/}.} 
Trevi,\footnote{\url{http://phototour.cs.washington.edu/patches/default.htm}.} 
and P53,\footnote{\url{http://archive.ics.uci.edu/ml/datasets/p53+Mutants}.} as well as two large-scale data sets Deep100M and Sift100M, where they contain the first $10^8$ data points that are respectively extracted from Deep1B \cite{babenko2016efficient} and ANN\_SIFT1B.\footnote{\url{http://corpus-texmex.irisa.fr/}.}
They cover a wide range of data types, including text, image, audio, biology, and rating data. 
We first remove the duplicate data points; then, we follow \cite{huang2021point} and randomly generate $100$ hyperplane queries for each data set. The statistics of the 16 data sets are summarized in Table \ref{table:dataset}. 

\subsection{Evaluation Metrics}
\label{sect:expt:metric}
\vspace{-0.1em}
We use the following metrics for performance evaluation. 
\begin{itemize}
\item \textbf{Indexing Time and Index Size} are estimated by the wall-clock time and memory usage of a method to build index, respectively. We use the indexing time and index size to evaluate the indexing overhead of a method. 

\item \textbf{Recall} is defined by the fraction of the total amount data points returned by a method that are appeared in the exact $k$ closest data points to the hyperplane query. We use recall to measure the accuracy of a method. 

\item \textbf{Query Time} is estimated by the wall-clock time of a method to answer the hyperplane query. We use this measure to evaluate the efficiency of a method. 

\end{itemize}

We run each method for each experiment five times to report its average recall, query time, and indexing overhead.

\subsection{Benchmark Methods}
\label{sect:expt:methods}
To the best of our knowledge, there is no proximity graph-based method for performing P2HNNS, and it is non-trivial to adapt them for solving P2HNNS as this problem is quite different from the classic similarity search problems. To make a fair comparison with \textbf{Ball-Tree} and \textbf{BC-Tree}, we choose two state-of-the-art hashing schemes, \textbf{NH} and \textbf{FH}, as baselines.\footnote{\url{https://github.com/HuangQiang/P2HNNS}.} 

For the problem of $k$-P2HNNS, we consider $k \in \{1,10,20,$ $40\}$. 
For Ball-Tree and BC-Tree, we set $N_0 \in \{ 100, 200, 500$, $1000, 2000, 5000, 10000\}$ and use the center preference by default. 
For NH and FH, we use their suggested versions with randomized sampling for asymmetric transformation, which can significantly reduce the indexing overhead while maintaining excellent query performance. 
Moreover, we follow \cite{huang2021point} to set up the parameters of NH and FH, i.e., we set the sampling dimension $\lambda \in \{d,2d,4d,8d\}$ and the hash table number $m \in \{8,16,32,64,128,256\}$ for both NH and FH and set the separation threshold $l \in \{2,4,6\}$ for FH; for the remaining parameters, we use their default values \cite{huang2021point}. 

\begin{figure*}[t]
\centering
\includegraphics[width=0.99\textwidth]{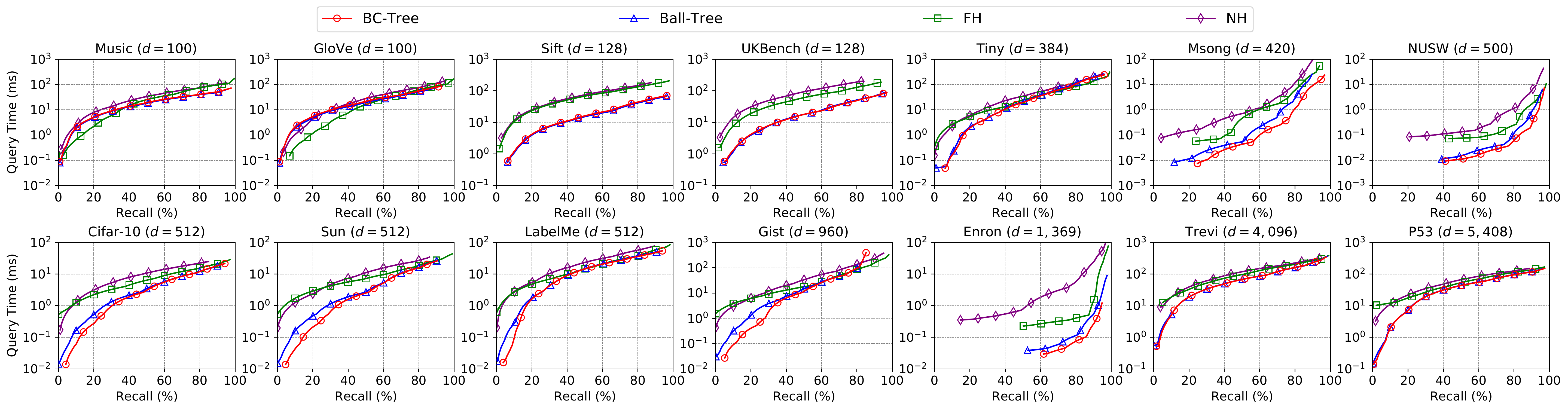}
\vspace{-1.0em}
\caption{Query time-recall curves of retrieving top-10 results.}
\label{fig:time-recall}
\vspace{-1.0em}
\end{figure*}


\begin{figure*}[t]
\centering
\includegraphics[width=0.99\textwidth]{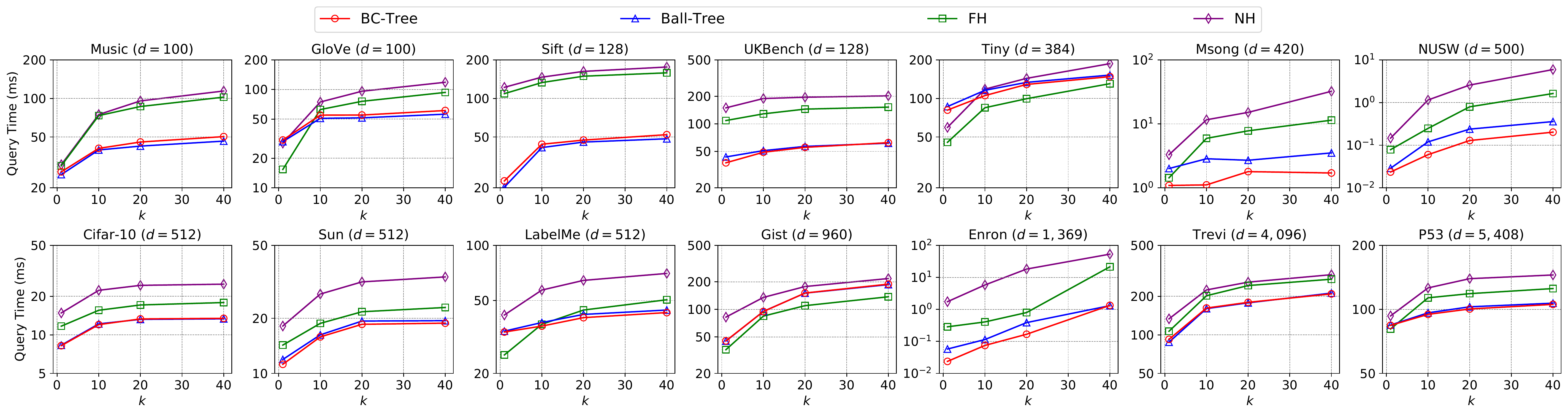}
\vspace{-1.0em}
\caption{Query time-$k$ curves at about 80\% recall.}
\label{fig:time-k}
\vspace{-1.5em}
\end{figure*}

\subsection{Indexing Performance}
\label{sect:expt:index}
We first study the indexing performance of Ball-Tree and BC-Tree. To make a fair comparison, we report the indexing overhead of NH and FH with $m=128$ as their query results with $m < 128$ are unreliable and unstable. To make a trade-off, we show the indexing performance of Ball-Tree and BC-Tree with $N_0=100$, leading to the largest space and time costs as the tree is the highest. Their results are listed in Table \ref{tab:index}.

The indexing time of Ball-Tree and BC-Tree is around 1.5$\sim$ 170$\times$ less than that of NH and FH. This can be explained by their indexing time complexities. 
According to Theorems \ref{theorem:ball_tree_construction} and \ref{theorem:bc_tree_construction}, the construction time of Ball-Tree and BC-Tree is $\tilde{O}(nd)$, while NH needs $O(n^{1+\rho}\lambda)$ time, where $\rho$ ($0<\rho<1$) is the LSH performance indicator. Similar to Ball-Tree and BC-Tree, FH takes $\tilde{O}(n\lambda)$ time to build hash tables, but it requires much extra cost for data partitioning \cite{huang2021point}. Note that we only consider the sampling dimension $\lambda$ in this experiment. if without randomized sampling, as $\lambda \rightarrow \Omega(d^2)$, the indexing time of NH and FH will be significantly longer. 

As for the index size, the advantage of Ball-Tree and BC-Tree is more apparent. Their index size is about 11$\sim$2,400$\times$ smaller than that of NH and FH. This is because Ball-Tree and BC-Tree only need $O(nd)$ space to store the centers in their nodes (in the worst case), while NH and FH require $O(n^{1+\rho})$ and $O(n\log n)$ space to store hash tables, respectively. Moreover, as FH uses a series of partitions for pruning, it needs extra space to store LSH functions for each partition. 

Compared with Ball-Tree, BC-Tree enjoys 1.0$\sim$1.2$\times$ less indexing time, which demonstrates the effectiveness of Lemma \ref{lemma:center_linear_property} to reduce the center computation cost for the internal nodes of BC-Tree.
However, BC-Tree also uses 1.01$\sim$1.57$\times$ larger index size as it requires extra $\Theta(n)$ space to store $r_{\bm{x}}$, $\norm{\bm{x}} \cos\varphi_{\bm{x}}$ and $\norm{\bm{x}} \cos\varphi_{\bm{x}}$ for each data point $\bm{x}$ for point-level pruning. 
Overall, their differences are not significant, which is consistent with our analysis in Theorem \ref{theorem:bc_tree_construction}. 
Moreover, the index size of Ball-Tree and BC-Tree is at least 11$\times$ smaller than the size of data sets shown in Table \ref{table:dataset} because we set the leaf size $N_0 = 100$, which is much larger than 1. Thus, the number of nodes they contain is less than $n$. 

\begin{figure*}[t]
\centering
\includegraphics[width=0.99\textwidth]{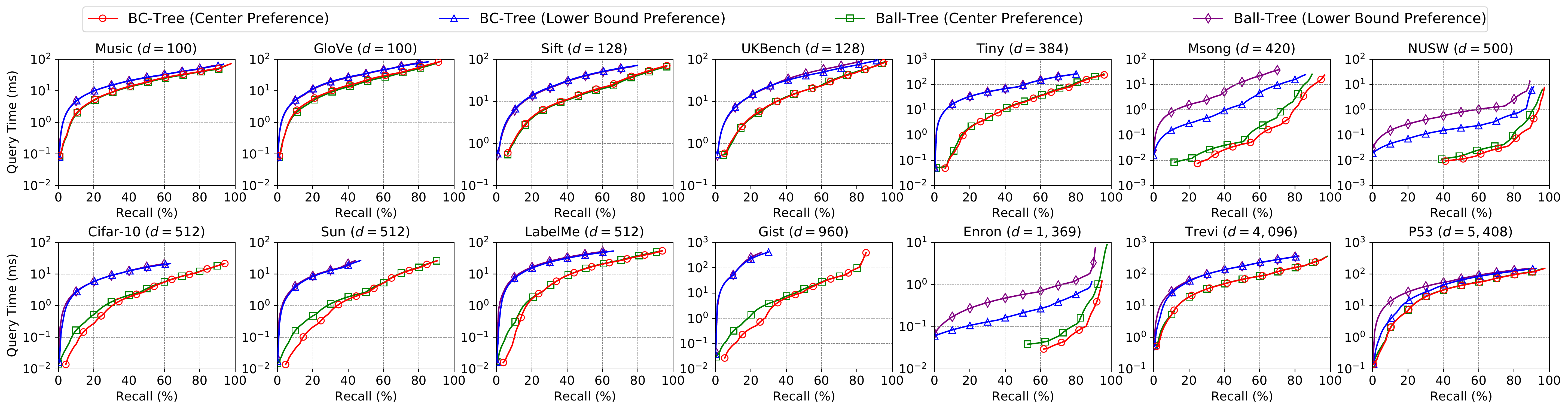}
\vspace{-1.0em}
\caption{The impact of the branch preference choice for BC-Tree and Ball-Tree.}
\label{fig:preference}
\vspace{-1.0em}
\end{figure*}

\begin{figure*}[t]
\centering
\includegraphics[width=0.99\textwidth]{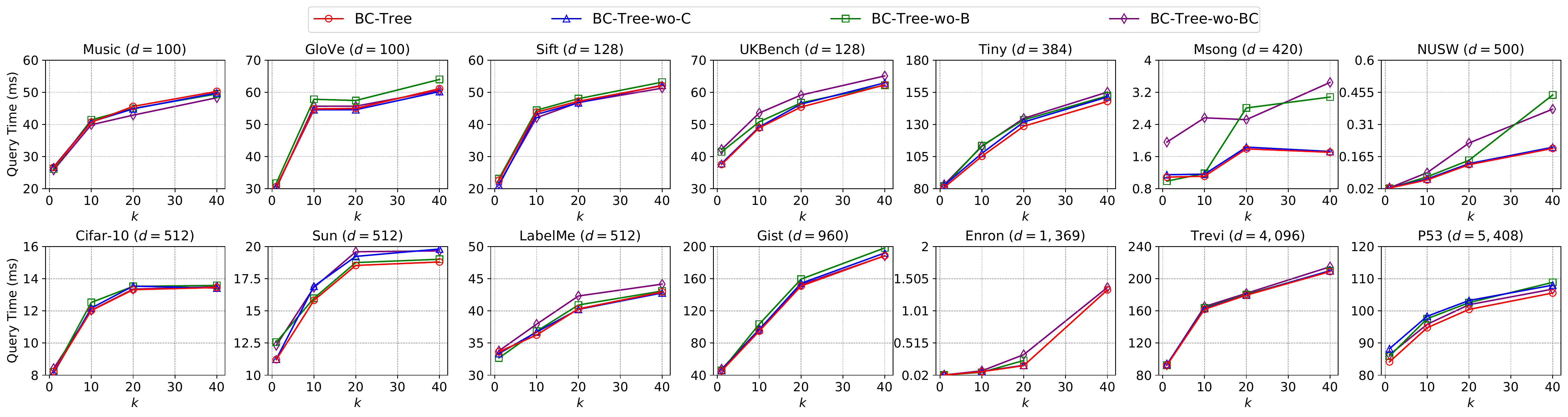}
\vspace{-1.0em}
\caption{The effectiveness of the individual lower bounds of BC-Tree.}
\label{fig:lower-bounds}
\vspace{-1.75em}
\end{figure*}

\subsection{Query Performance}
\label{sect:expt:query}
To justify the query performance of Ball-Tree and BC-Tree, we set different candidate fractions to achieve different recalls. To alleviate the impact of parameters, we report the lowest query time of a method for a certain recall from all its parameter combinations. The results with $k=10$ are shown in Figure \ref{fig:time-recall}. Similar trends can be observed in other $k$ values.

BC-Tree and Ball-Tree are about 1.1$\sim$10$\times$ faster than the better of NH and FH on 12 out of 14 data sets (except Tiny and Gist). The reasons are two folds. (1) The asymmetric transformation of NH and FH leads to a significant distortion error for performing NNS and FNS. Even though their query time complexity is sublinear to $n$, as it is hard to distinguish the close points to the query from the far ones, their practical performance is poor. (2) The lower bounds for Ball-Tree and BC-Tree are practical yet effective with the simple ball and cone structures. Thus, Ball-Tree and BC-Tree perform well, even on moderate- and high-dimensional data sets.

Moreover, the advantage of Ball-Tree and BC-Tree in terms of the query time (except Music and GloVe) is much more apparent when the recall is less than 60\%, especially BC-Tree. 
This might be because their traversing cost to reach the leaves is cheap. 
Further, the collaborative inner product computing strategy (Lemma \ref{lemma:ip_linear_property}) can reduce the lower bound computation cost for the internal nodes of BC-Tree. 
In contrast, the hashing-based methods require a higher cost to compute the hash functions before verifying candidates in the colliding buckets. 

From Figure \ref{fig:time-recall}, we also discover that BC-Tree is more efficient than Ball-Tree on 7 out of 14 data sets. These results justify the effectiveness of the point-level pruning and the collaborative inner product computing strategies. 
For the remaining 7 data sets, their performance is very close. The reason might be that these data sets are more complicated than the others, making the two strategies less effective. 

\subsection{Sensitivity to $k$}
\label{sect:expt:k}
We consider $k \in \{1,10,20,40\}$ and study the sensitivity of Ball-Tree and BC-Tree to $k$. We plot the query time-$k$ curves of all methods at about $80\%$ recall in Figure \ref{fig:time-k}. 

We observe that the query time-$k$ curves show similar trends to the query-time recall curves, which further validate the superior query performance of Ball-Tree and BC-Tree. 
Moreover, similar to NH and FH, the query time of Ball-Tree and BC-Tree increases a lot for $k$ from 1 to 10, but as $k$ continues to increase, all of them become less sensitive to $k$. 

\begin{figure}[t]
\centering
\includegraphics[width=0.37\textwidth]{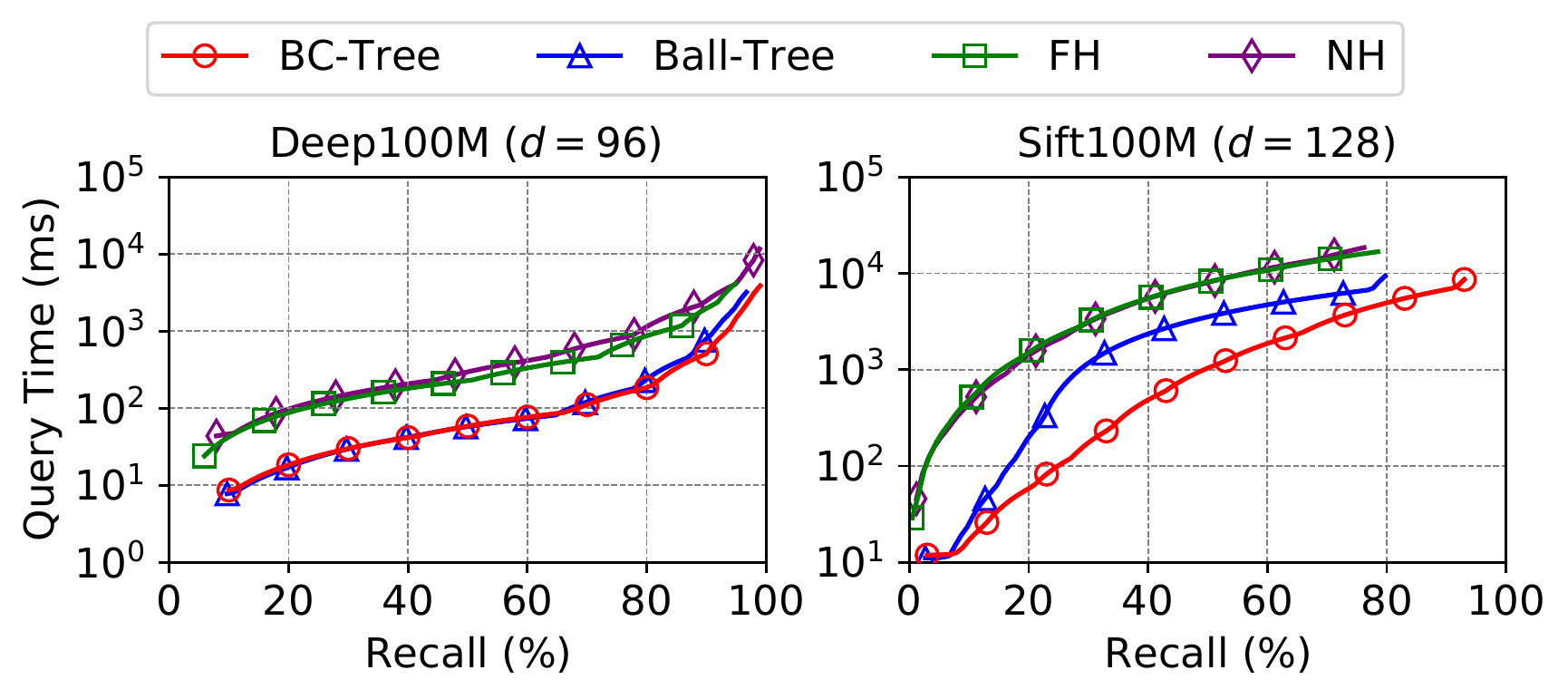}
\vspace{-1.25em}
\caption{The query performance on Deep100M and Sift100M ($k=10$).}
\label{fig:scalability}
\vspace{-1.25em}
\end{figure}

\subsection{Branch Preference Choice}
\label{sect:expt:preference}
We then study the impact of the branch preference choice for Ball-Tree and BC-Tree. The query time-recall curves of Ball-Tree and BC-Tree with the center preference and lower bound preference are depicted in Figure \ref{fig:preference}. 

Ball-Tree and BC-Tree with the center preference are around 2$\sim$100$\times$ faster than those with the lower bound preference, especially when the recall is less than 60\%. These results demonstrate that the center preference is uniformly better than the lower bound preference for the P2HNNS, which is in accordance with our analysis discussed in Section \ref{sect:ball:search}. 

\subsection{Effectiveness of Individual Lower Bounds of BC-Tree}
\label{sect:expt:lower_bound}
We evaluate the effectiveness of the individual lower bounds of BC-Tree. We plot the query time-$k$ curves of different variants of BC-Tree at about 80\% recall in Figure \ref{fig:lower-bounds}, 
where BC-Tree-wo-C, BC-Tree-wo-B, and BC-Tree-wo-BC represent the BC-Tree without the point-level cone bound, point-level ball bound, and both point-level bounds, respectively.

We have three interesting discoverings. 
(1) Both BC-Tree-wo-C and BC-Tree-wo-B enjoy less query time than BC-Tree-wo-BC on most data sets, which validates that the point-level ball bound and the point-level cone bound are effective for pruning false positive data points. 
(2) BC-Tree is the fastest method among the four competitors. This discovery indicates that combining the point-level ball bound and the point-level cone bound can lead to the best pruning effectiveness. 
(3) BC-Tree-wo-C is faster than BC-Tree-wo-B on most data sets. This is because although the point-level cone bound is tighter than the point-level ball bound, it is more complicated than the point-level ball bound, leading to a higher computation cost. 

\begin{figure}[t]
\centering
\includegraphics[width=0.38\textwidth]{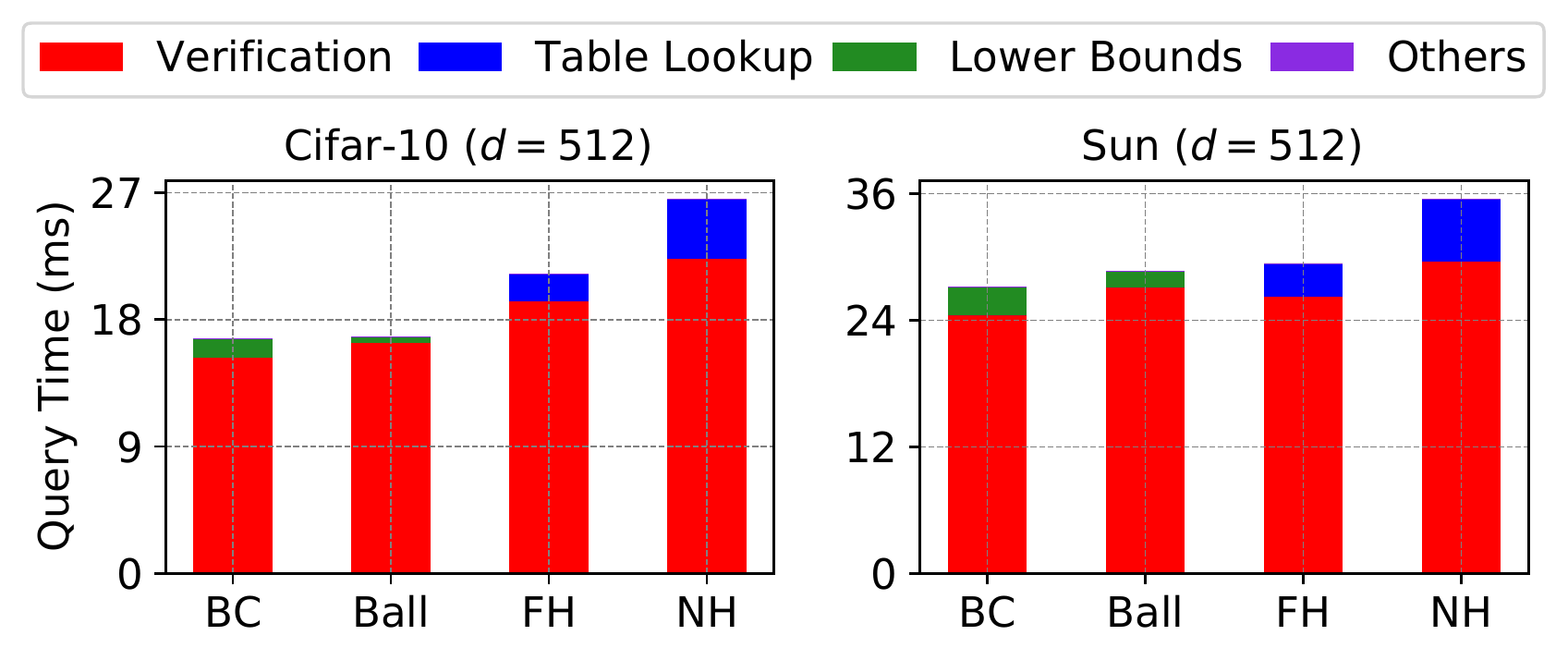}
\vspace{-1.0em}
\caption{Time profile visualization on Cifar-10 and Sun.}
\label{fig:time_profile}
\vspace{-1.5em}
\end{figure}

\begin{figure*}[t]
\centering
\includegraphics[width=0.99\textwidth]{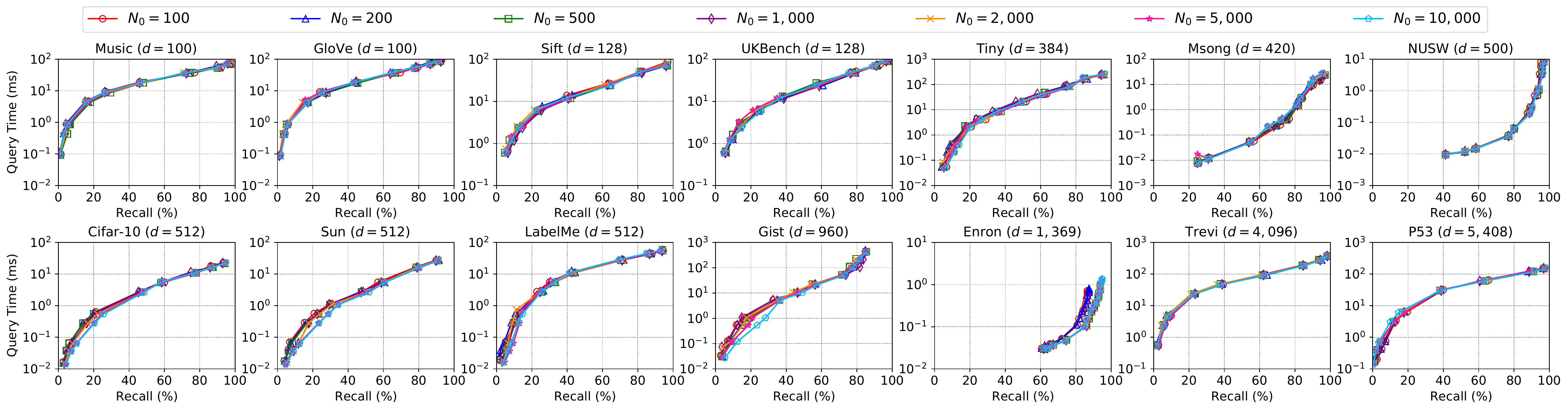}
\vspace{-1.0em}
\caption{The impact of the leaf size $N_0$ of BC-Tree.}
\label{fig:leaf_size}
\vspace{-1.5em}
\end{figure*}

\subsection{Performance on Large-Scale Data Sets}
\label{sect:expt:scalability}
We then justify the scalability of Ball-Tree and BC-Tree on two large-scale data sets Sift100M and Deep100M. 
We show the query time-recall curves with $k=10$ in Figure \ref{fig:scalability} and depict their indexing time and index size in Table \ref{tab:index}. 

The indexing and query performance trends of the four methods on Sift100M and Deep100M are similar to those on 14 small data sets. Moreover, BC-Tree shows higher efficiency superiority than FH and NH on Sift100M and Deep100M, especially the query time, e.g., BC-Tree is at least 10$\times$ faster than FH and NH on Sift100M for the recall in $[20\%, 40\%]$. The speedup ratio is the largest among the 16 data sets.

\subsection{Time Profile Visualization}
\label{sect:expt:profile}
We visualize the time profile of the four methods to illustrate where the time they spend. The results at about 90\% recall on Cifar-10 and Sun are shown in Figure \ref{fig:time_profile}.

To reach 90\% recall, all four methods spend the most time on candidate verification. 
The table lookup time of FH and NH is uniformly larger than the lower bound computation time of Ball-Tree and BC-Tree,  which further explains the superior efficiency of Ball-Tree and BC-Tree when the recall is less than 60\% (Section \ref{sect:expt:query}). 
Besides, even though BC-Tree spends more time on lower bound computation than Ball-Tree, its total query time is smaller, which further validates the pruning effectiveness of the two point-level lower bounds.

\subsection{Impact of the Leaf Size $N_0$ of BC-Tree}
\label{sect:expt:leaf}
Finally, we study the impact of the leaf size $N_0$ for BC-Tree to guide the users for parameter setting. The query time-recall curves of BC-Tree are depicted in Figure \ref{fig:leaf_size}. 

We have two interesting observations.
First, the query time of BC-Tree under different $N_0$ is close, especially for the recall $>80\%$.
The reasons might be that (1) the collaborative inner product computing strategy can reduce the total lower bound computation cost by almost half when traversing the tree; (2) the two point-level lower bounds can avoid the exhaustive scan and reduce the total candidate verification cost. Thus, BC-Tree is not very sensitive to $N_0$. 
Second, the query time of BC-Tree with $N_0=100$ and $200$ on Gist and Enron is larger than that with other settings, which means that a larger $N_0$ might be more satisfied for high-dimensional data sets. 

\section{Related work}
\label{sect:related_work}
\vspace{-0.5em}
\paragraph{P2HNNS}
The problem of P2HNNS has received increasing attention. We first review the pioneer works.
The first sublinear time method to tackle P2HNNS was proposed by Jain et al. \cite{jain2010hashing} in 2010. They introduced two hyperplane hashing schemes, AH and EH, that are locality-sensitive to the angle between the data points and the vector normal to the hyperplane query. 
Later, BH \cite{liu2012compact} and MH \cite{liu2016multilinear} were developed to boost the query performance of AH and EH, which aim to use more linear hash functions to amplify the difference in the collision probabilities. 
Recently, with the asymmetry design of hash functions, Aum{\"u}ller et al. \cite{aumuller2018distance} presented a general distance-sensitive hashing scheme beyond LSH. 
The above hyperplane hashing schemes, however, only conditionally deal with the P2HNNS. They are only effective for normalized data. 
Recently, Huang et al. \cite{huang2021point} designed the first two sublinear time hashing schemes NH and FH, which directly solve the P2HNNS beyond the unit hypersphere. Nevertheless, their asymmetric transformations lead to a considerable overhead in constructing hash tables and a vast distortion error. 

Until now, all pioneer works have focused on hashing-based methods. 
Although tree-based methods have been well studied in many similarity search tasks, too little work has been devoted to utilizing them for P2HNNS. 
This paper revisits a classical Ball-Tree index. The proposed Ball-Tree and BC-Tree demonstrate superior performance against NH and FH. 

\paragraph{MIPS}
The MIPS problem aims to find the data with the largest inner product value to a given query, while the P2HNNS can be regarded as a minimum absolute inner product search problem. 
The two problems share similar nature, i.e., they require computing the inner product, and their distance/similarity functions are not metric. 
There exist many representative works that apply tree structures to tackle MIPS, such as Ball-Tree \cite{ram2012maximum}, Metric-Tree \cite{koenigstein2012efficient}, and Cover-Tree \cite{curtin2013fast,curtin2014dual}. 
Though these methods are non-trivial to be adapted for performing P2HNNS as their aims are quite different and the problem of P2HNNS contains an extra absolute value operator, they motivate this work and inspire us to design the new, tight lower bounds for Ball-Tree and BC-Tree for solving P2HNNS.

\section{Conclusions}
\label{sect:conclusions}
In this paper, we study a new yet very challenging problem of P2HNNS. We start with investigating a vanilla Ball-Tree index and propose a simple branch-and-bound method with a novel lower bound.  
Then, we build upon the Ball-Tree and design a new tree structure named BC-Tree.
BC-Tree inherits both the lightweight and inexpensive construction cost of Ball-Tree while providing a similar or more efficient hyperplane query response. 
Extensive experiments over 16 real-world data sets confirm their superior indexing and query performance. 
The excellent performance of Ball-Tree and BC-Tree beyond the hashing-based methods might shed a light on revitalizing tree-based methods for similarity search. 

\section*{Acknowledgment}
We sincerely thank Dr. Jianlin Feng and Dr. Yikai Zhang for their valuable discussions in the earlier stages of this work. 
This research is supported by the National Research Foundation, Singapore under its Strategic Capability Research Centres Funding Initiative. Any opinions, findings and conclusions or recommendations expressed in this material are those of the author(s) and do not reflect the views of National Research Foundation, Singapore.

\balance

\bibliographystyle{IEEEtranS}
\bibliography{IEEEabrv,main}

\end{document}